\documentclass[journal]{IEEEtran}
\normalsize

\usepackage{cite}
\usepackage{graphicx}
\usepackage[cmex10]{amsmath}
\usepackage{amssymb}
\usepackage{booktabs}
\usepackage{threeparttable}
\usepackage{algorithm}
\usepackage{algpseudocode}
\usepackage{amsfonts}
\usepackage{xcolor}
\usepackage{color}

\usepackage{tabularx} %

\usepackage{dblfloatfix} %

\usepackage{soul}

\usepackage{cancel} %

\usepackage{verbatim}%

\usepackage{makecell}

\usepackage{empheq} %

\usepackage{underoverlap} %

\usepackage{setspace}

\usepackage{hyperref}
\hypersetup{pdftex,colorlinks=true,allcolors=black}

\usepackage{calc}

\usepackage{textcomp}
\def\BibTeX{{\rm B\kern-.05em{\sc i\kern-.025em b}\kern-.08em
		T\kern-.1667em\lower.7ex\hbox{E}\kern-.125emX}}

\newtheorem{remark}{\bf  Remark}

\newtheorem{proposition}{\bf Proposition}
\newtheorem{corollary}{\bf  Corollary}

\newcommand{\HF}{hopping frequency}
\newcommand{\HFs}{hopping frequencies}

\newcommand{\mj}{\mathsf{j}}

\begin{document}

\title{
\parbox{\textwidth}{\centering Waveform Design and Accurate Channel Estimation for Frequency-Hopping MIMO Radar-Based Communications}
}

\author{
	Kai Wu,
	 J. Andrew Zhang,~\IEEEmembership{Senior Member,~IEEE}, %
	Xiaojing Huang,~\IEEEmembership{Senior Member,~IEEE}, \\
	Y. Jay Guo,~\IEEEmembership{Fellow,~IEEE}, and Robert W. Heath Jr.,~\IEEEmembership{Fellow,~IEEE}
	 \vspace{-20pt}
	 \thanks{K. Wu, J. A. Zhang, X. Huang and Y. J. Guo are with the Global Big Data Technologies Centre, University of Technology Sydney, Sydney, NSW 2007, Australia (e-mail: \{kai.wu; andrew.zhang; xiaojing.huang; jay.guo\}@uts.edu.au).}%
	\thanks{R. W. Heath Jr. is with the University of Texas at
		Austin, Austin, TX 78712 USA (e-mail: rheath@utexas.edu).}
}


\maketitle

\begin{abstract}
	
	Frequency-hopping (FH) MIMO radar-based dual-function radar communication (FH-MIMO DFRC) enables communication symbol rate to exceed radar pulse repetition frequency, which requires accurate estimations of timing offset and channel parameters.
	The estimations, however, are challenging due to unknown, fast-changing \HFs~and the multiplicative coupling between timing offset and channel parameters. 
	In this paper, we develop accurate methods for a single-antenna communication receiver to estimate timing offset and channel for FH-MIMO DFRC. 
	\textit{First}, we design a novel FH-MIMO radar waveform, which enables a communication receiver to estimate the \HF~sequence (HFS) used by radar, instead of acquiring it from radar.
Importantly, the novel waveform incurs no degradation to radar ranging performance. 
	\textit{Then}, via capturing distinct HFS features, we develop two estimators for timing offset and derive mean squared error lower bound of each estimator. Using the bounds, we design an HFS that renders both estimators applicable.
	\textit{Furthermore}, we develop an accurate channel estimation method, reusing the single hop for timing offset estimation.
	{Validated by simulations, the accurate channel estimates attained by the proposed methods enable the communication performance of DFRC to approach that achieved based on perfect timing and ideal knowledge of channel.}
\end{abstract}


\section{Introduction}\label{sec: intro}

There have been increasing demands for systems with
both communications and radar sensing capabilities on
emerging platforms such as unmanned aerial vehicles and smart
cars \cite{choi2016millimeter}. Instead of having two separate systems, it is possible to develop
techniques to integrate the two functions into one by sharing
hardware and signal processing modules, and achieve immediate
benefits of reduced cost, size, weight, and better spectrum
efficiency \cite{Andrew_Multibeam2019TVT}.
The design of joint radar and communication systems can be categorized into three groups: coexistence, cooperation and co-design \cite{Bliss_3catogeries2017}.
	In the first two groups,  the joint design, though improves the spectral efficiency,
	 has an inevitable issue of mutual interference between radar and communications
\cite{aydogdu2019radchat}.	 
In co-design, dual-function waveform can be optimized by jointly considering radar and communication performance metrics (e.g., mutual information and achievable rate) \cite{sturm2011waveformJCAS,Bliss_survey2017,LiuFan_waveform2018TSP}, which, however, can have constrained radar sensing ability, as compared with using dedicated radar waveform.

Conducting data communications based on radar platforms, referred to as dual-function radar communication (DFRC), has been studied for decades \cite{DFRC_SP_Mag2019Amin_Aboutanios}. Given the popularity of the frequency modulated continuous wave (FMCW) radars particularly in automotive applications, early DFRC tends to consider FMCW radars \cite{barrenechea2007fmcw}. 
	The advancement of MIMO radars has made the recent DFRC designs in favor of considering this modern radar system \cite{DFRC_SidelobeControl2016TSP,DFRC_SparseArray2019TAES_XianrongWang,DFRC_waveformShuffling2018DSP,DFRC_CSK2018DSP}.
Some researchers optimize the beam pattern of a MIMO radar to perform conventional modulations, such as phase shift keying (PSK) and amplitude shift keying, using sidelobes in the MIMO radiation patterns \cite{DFRC_SidelobeControl2016TSP,DFRC_SparseArray2019TAES_XianrongWang}. Others optimize radar waveform to perform non-traditional modulations, such as waveform shuffling \cite{DFRC_waveformShuffling2018DSP} and code shift keying \cite{DFRC_CSK2018DSP}. These works \cite{DFRC_SidelobeControl2016TSP,DFRC_SparseArray2019TAES_XianrongWang,DFRC_waveformShuffling2018DSP,DFRC_CSK2018DSP} embed one symbol per one or multiple radar pulses; hence the communication symbol rate is limited by radar pulse repetition frequency (PRF).

Employing frequency-hopping (FH) based MIMO (FH-MIMO) radar can increase the symbol rate to much larger than radar PRF, since each radar pulse is divided into multiple sub-pulses (also referred to as hops) and information embedding can be performed on a sub-pulse basis  \cite{DFRC_AmbiguityFunc2018Amin,FH_MIMO_Radar2019_RadarConf,DFRC_FHcodeSel2018}. 
For brevity, we refer to FH-MIMO radar-based DFRC as FH-MIMO DFRC. There are two schemes of FH-MIMO DFRC {that embed one communication symbol per hop}. The first scheme
embeds a PSK symbol into radar signal per antenna and hop \cite{DFRC_AmbiguityFunc2018Amin,FH_MIMO_Radar2019_RadarConf}.  
To demodulate PSK symbols, a communication receiver {needs an accurate estimate of the channel response}.
The second realization exploits different combinations of hopping frequencies as constellation symbols, referred to as FH code selection (FHCS) \cite{DFRC_FHcodeSel2018}, but does not necessarily require channel estimation. The achievable rate of FHCS can be limited by the number of different combinations of hopping frequencies, which, nevertheless, can be improved by combining PSK. The combination, as with sole PSK, requires an accurate channel estimate for decoding.

An accurate channel estimation plays an important role in the FH-MIMO DFRC. 
Channel estimation, however, is challenging for various reasons.
{First, training signals for channel estimation can incur changes to existing radar waveforms, which, possibly, results in undesirable performance degradation to radar detection.}
Second, the pairing between hopping frequencies and antennas is a critical information in channel estimation (as will be clear in Section \ref{subsec: problem formulation}) and information decoding \cite{FH_MIMO_Radar2019_RadarConf,DFRC_FHcodeSel2018}; however, the acquisition of the pairing information at a communication receiver is non-trivial and depends on how much information radar shares.
Third, fine timing, which finds a precise timing offset value, is not easy to realize in FH-MIMO DFRC (although a coarse timing can be achieved at a communication receiver by performing conventional energy-based/auto-correlation packet detection \cite{DFRC_ChnnlEst_FanLiu_howMuchInfo}).

\subsection{Main Contributions of Our Work}
{Aiming to tackle the above challenges, we discover and exploit the unique waveform structure of FH-MIMO radar to	
	 develop low-complexity and high-accuracy estimation methods for timing offset and channel parameters. The main contributions are summarized as follows.}
	 
	 \begin{enumerate}
	 	\item {We design a novel FH-MIMO radar waveform by introducing a simple re-ordering processing per hop. Enabled by the novel waveform, we propose to estimate, instead of acquiring from radar, the \HF~used by each radar transmitter antenna per hop. Importantly, we also prove that the novel waveform incurs no degradation to the radar ranging performance;}

	 	\item {For line-of-sight (LoS) channels, we develop two timing offset estimators which are suitable for distinct \HF~sequences and levels of communication signal-to-noise ratio (SNR). We also derive mean squared error (MSE) lower bound (MSELB) of each estimator. Using the bounds, we further design a sub-optimal \HF~sequence which renders both estimators applicable. With the timing offset estimated, we develop methods to accurately estimate the remaining channel parameters, applying our recent work \cite{Kai_freqEst2020CL};}
	 	
	 	\item {For multi-path channels, we develop a method to estimate the composite of multiple paths using incomplete sampled hops, where we 
	 	 propose to combat inter-hop and inter-antenna interference by judiciously configuring hopping frequency. 
	 	 Then, we extend the methods developed for LoS channel to multi-path channels after recovering the pivotal phase information using the estimated multi-path composite. 
 	 }

	 \end{enumerate}

\noindent 
To improve the communication data rate in FH-MIMO DFRCs, we propose to combine PSK \cite{FH_MIMO_Radar2019_RadarConf} and FHCS \cite{DFRC_FHcodeSel2018}, referred to as PFHCS. We also provide a PFHCS demodulation scheme applying the estimated timing offset and channel parameters.
Simulations are provided to validate the high accuracy of the proposed methods and the improved communication performance of PFHCS compared with PSK \cite{FH_MIMO_Radar2019_RadarConf} and FHCS \cite{DFRC_FHcodeSel2018}.
In particular, the timing offset estimation is able to approach the derived MSELB across a wide SNR region. At a moderate SNR of $ 15 $ dB, our schemes enable data rate and symbol error rate (SER) to approach those based on perfect timing offset and ideal knowledge of channel parameters.

	\subsection{Literature Review}	
\subsubsection{Channel Estimation Methods for DFRC}
To the best of our knowledge, there has been no published work on channel estimation methods for FH-MIMO DFRC. As a matter of fact, only a few works \cite{DFRC_ChnnlEst_sparse2016,DFRC_ChnnlEst_sparse2017} develop channel estimation methods for radar-based communications. Specifically, sparse recovery-based channel estimation methods are developed in \cite{DFRC_ChnnlEst_sparse2016,DFRC_ChnnlEst_sparse2017} which coordinate radar and communication receiver using probing beams.
In a different yet relevant context (spectrum sharing), interference channel between radar and communication is estimated to achieve co-existence  \cite{DFRC_ChnnlEst_interferenceCoexistence2013,DFRC_ChnnlEst_controlCentre,DFRC_ChnnlEst_FanLiu_howMuchInfo}.
	In \cite{DFRC_ChnnlEst_interferenceCoexistence2013}, communication users are coordinated to send training symbols and radar performs the maximum-likelihood (ML) estimation on communication channels. In \cite{DFRC_ChnnlEst_controlCentre}, radar and communication are scheduled by a control center.  Different from the above works, uncoordinated radar and communication base station (BS) are considered in \cite{DFRC_ChnnlEst_FanLiu_howMuchInfo}, where
	several hypothesis testing and ML estimators are developed for channel estimation.   

In some recent DFRC works \cite{Andrew_Multibeam2019TVT,Andrew_perspectiveJCAS2020TAES}, 
channel estimation methods are developed, which, however, are based on new (future) DFRC waveforms/platforms \cite{Andrew_OptMultibeam2019TCom}. 
	These new designs are specifically tailored for dual functions and are non-trivial to be applied in FH-MIMO DFRC.
A common feature captured by most of the above methods is the full cooperation between radar and communication. 
	An exception is found in \cite{DFRC_ChnnlEst_FanLiu_howMuchInfo} which, as in the present work, considers an uncooperative scenario where no communication-specific training signal is available from radar. 
	Unlike \cite{DFRC_ChnnlEst_FanLiu_howMuchInfo} which is based on a conventional MIMO radar and a BS potentially with strong computing power, we consider an FH-MIMO radar and a low-profile communication receiver with low computing power and few (or a single) antennas.

	\subsubsection{Information Embedding Schemes for FH-MIMO DFRC}\label{subsubsec: information embedding schemes}

	Aimed at improving radar spectral efficiency and reducing range sidelobe levels, 
	the differential PSK (DPSK) \cite{DFRC_DPSK2019Amin} and continuous phase modulation (CPM) \cite{DFRC_cskCPM} have also been considered in FH-MIMO DFRC.
	In \cite{DFRC_pskDPSKcpm2020}, a comprehensive analysis is provided to compare PSK, DPSK and CPM in terms of their impact on radar ranging and their data rate. 
	From the communication perspective, it is known that the asymptotic SER performance of DPSK is generally worse than PSK, particularly when the modulation order is high; and the optimal receiver of CPM can be complicated to implement \cite{SER_DPSKvsPSK}. 
	Most of the above FH-MIMO DFRC schemes require channel information for communication decoding. Although the information is not necessarily required by FHCS and DPSK, the timing offset has to be estimated and compensated to avoid inter-hop interference. Neither channel nor timing offset estimation is considered in these works.  
	For ease of exposition, we employ PSK and FHCS to develop timing offset and channel estimation methods, since they have simpler signal models than the other modulations. 
	The proposed methods can be readily applied for other modulations by using similar signal frame structure, as to be designed in Section \ref{subsec: proposed frame structure}.

\subsection{Paper Structure and Notations}
The remainder of the paper is organized as follows. In Section \ref{sec: signal model}, the FH-MIMO radar is described first, then PSK \cite{FH_MIMO_Radar2019_RadarConf} and FHCS \cite{DFRC_FHcodeSel2018} are briefly reviewed, and the problems of estimating timing offset and channel parameters are also formulated. 
In Section \ref{sec: new waveform}, a novel FH-MIMO radar waveform is designed, and accordingly the overall channel estimation scheme is developed. Section \ref{sec: tau estimaiton} first develops two estimators for timing offset and then analyzes their performance, leading to the design of a sub-optimal \HF~sequence.
The remaining channel parameters are estimated in Section \ref{sec: estimate u and beta and perform communciations}, followed by the design of PFHCS and its demodulation in Section \ref{sec: decoding  using estimated channels}. 
Extension of the proposed methods to multi-path and multi-antenna scenarios is elaborated on in Section \ref{sec: extensions to NLOS and multi-antenna scenarios}. 
Simulation results are provided in Section \ref{sec: simulations} with conclusions provided in Section \ref{sec: conclusion}.

	\textit{Notations:} The following notations/rules are followed. $ C_M^K $ denotes binomial coefficient. $ \lfloor\cdot\rceil $ rounds towards nearest integer. $ \Re\{x\} $ takes the real part of $ x $ and $ \Im\{\} $ the imaginary part.
	$ |\cdot| $ can take amplitude, absolute and cardinality, depending on context. $ \mathcal{G}\{\} $ denotes the greatest common divisor (GCD). 
	$ \angle x $ takes the angle of $ x $. 
	$ \hat{x} $ denotes the estimate of $ x $ and $ \measuredangle x $ the angle estimate. 	$ \mathbb{I}_{+} $ denotes the set of positive integers, $ \mathbb{N} $ the set of nature numbers and $ \mathbb{C} $ the set of complex numbers.
	Variables with subscripts $ ()_{h} $ and $ ()_{m} $ indicate their associations with hop $ h $ and antenna $ m $, respectively.
	Table \ref{tab: notation list} summarizes the notations used in this paper.

\begin{table}[!t]\footnotesize
	\caption{{Notations and Definitions}}
	\begin{center}
		\begin{tabularx}{85mm}{l X}
			\toprule[0.5pt]
			$ H $    & number of hops per radar pulse \\
			$ K $    & number of radar sub-bands in frequency band $ B $ \\		
			$ L(=T/T_{{\mathrm{s}}}) $    & number of samples per radar hop \\
			$ L_{\eta}(=\eta/T_{{\mathrm{s}}}) $    & number of samples corresponds to $ \eta $ \\
			$ M $    & number of antennas of radar transmitter array \\
			$ T $   & hop duration \\	
			$ T_{\mathrm{s}} $    & sampling interval \\
			
			$ Y_h(l) $    & $ L $-point DFT of $ y_h(i) $ \\
			
			$ Y_m $ & $ Y_h(l^*_{M-1-m}) $, solely related to antenna $ m $; see (\ref{eq: DFT of com signal m-th frequency multipath})\\

			$ f_{\mathrm{L}} $    & lower limit of radar RF \\
			$ f_{hm} $    & hopping frequency used by antenna $ m $ at hop $ h $ \\
			$ f_k $    & frequency of sub-band $ k $ \\
			$ k_{hm} $    & index of the sub-band used by antenna $ m $ at hop $ h $  \\
			$ l_m^* $    & index of the discrete frequency of peak $ m $ of $ |Y_h(l)| $ \\
			$ y_h(t)\big/y_h(i) $    & continuous$ \big/ $digital signal received at hop $ h $ \\
			
			$ \eta $    & timing offset \\
			$ \phi $    & AoD from radar to communication end \\
			$ \beta $    & channel gain \\
			$ \xi(t)\big/\xi(i) $    & continuous$ \big/ $digital AWGN in the time domain \\
			$ \Xi(l) $    & $ L $-point DFT of $ \xi(i) $ \\
			$ \omega_{{\eta}},u,\tilde{\beta} $    & intermediate variables related to $ \eta $, $ \phi $ and $ \beta $, respectively; see (\ref{eq: intermediate variables}) \\	
			$ \varpi_{hm} $ & PSK modulation phase for antenna $ m $ at hop $ h $\\
			$ \kappa_{m} $ & second-order difference of $ \hat{k}_m $; see (\ref{eq: barY})\\
			
		$ \mathcal{M};\bar{\mathcal{M} };\breve{\mathcal{M} } $ & set of antenna indices satisfying $  \kappa_{m}\ne 0$$ ; $ $  |\kappa_{m}|=1$$ ; $ $  |\kappa_{m}|> 1$\\					
		$ \bar{{M} }; \breve{{M} } $ & cardinality of $ \bar{\mathcal{M}} $$ ; $ $ \breve{\mathcal{M}} $\\
			\toprule[0.5pt]
		\end{tabularx}
		\vspace{-20pt}
	\end{center}
	\label{tab: notation list}
\end{table}

	\section{Signal Model and Problem Formulation}\label{sec: signal model}
	
\begin{figure}[!t]
	\centerline{\includegraphics[width=85mm]{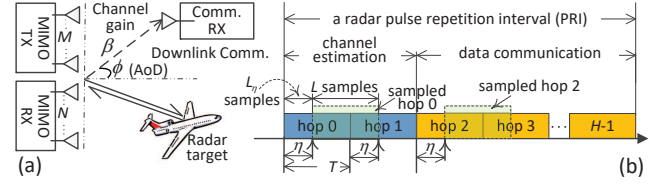}}%
	\caption{{(a) Illustration on the system diagram of an FH-MIMO DFRC; (b) The signal frame structure for the downlink communication in Fig. \ref{fig: signal structure}(a). The aircraft symbol is downloaded from \url{https://www.clipartkey.com/view/ThboRm_airplane-aircraft-vector-illustrator-flying-clipart-airplane-gif/}
	}}
	\label{fig: signal structure}
	\vspace{-10pt}
\end{figure}

	{Consider the FH-MIMO DFRC illustrated in Fig. \ref{fig: signal structure}(a). There are an FH-MIMO radar and a single-antenna user terminal. In addition to target illumination, the radar also performs downlink communication with a communication user through an LoS channel\footnote{{Note that LoS channel is generally considered in ground-to-air communications \cite{Channel_UAVchannel_zeng2019accessing}. 
				Therefore, our proposed methods are promising to be applied in an air-surveillance radar-based communications with aircrafts, where the radar is typically located in high altitudes \cite{DFRC_ChnnlEst_FanLiu_howMuchInfo}
				and aircrafts can be hundreds to thousands of meters above the sea level \cite{Channel_UAVchannel_zeng2019accessing}. 
		}}.} In this section, {we describe the signal model of the FH-MIMO radar}, based on which PSK \cite{DFRC_AmbiguityFunc2018Amin} and FHCS \cite{DFRC_FHcodeSel2018} are reviewed.

	\subsection{FH-MIMO Radar}\label{subsec: frequency hopping radar}
	The FH-MIMO radar of interest is based on fast frequency hopping. {Each pulse is divided into $ H $ sub-pulses}, i.e., \textit{hops} \cite{DFRC_FHcodeSel2018}. 
	The centroid frequency of the transmitted signal changes randomly across hops and antennas. 
	Denote the starting radio frequency (RF) of the radar as $ f_{\mathrm{L}} $ and the bandwidth as $ B $. By dividing the frequency band evenly into $ K $ sub-bands, the centroid frequency of the $ k $-th sub-band is given by 
	$ f_{k} = f_{\mathrm{L}}+\frac{kB}{K} ~(k=0,1,\cdots,K-1)$.
	Let $ M $ denote the number of antennas in the radar transmitter array. 
	Out of the $ K $ centroid frequencies, $ M(<K) $
	frequencies are selected to be the \HFs~at a hop, one per antenna. 
	{Denote the \HF~at hop $ h $ and antenna $ m $ as $ f_{hm} $ that satisfies $ f_{hm}\in \{f_k~\forall k\} $.}
	To ensure the waveform orthogonality of the FH-MIMO radar,
	the following are required  \cite{DFRC_AmbiguityFunc2018Amin,Amb_FH_MIMO2008TSP}
	\begin{align}\label{eq: fphm ne fphm'}
	f_{hm}\ne f_{hm'}~(\forall m\ne m', ~\forall h);~~{BT}/{K}\in \mathbb{I}_{+},
	\end{align}
	where $ T $ is hop duration. At hop $ h $ the $ m $-th antenna of the radar transmitter transmits
	\begin{align}\label{eq: transmitted signal p h m}
	{s}_{hm}(t) =	e^{-\mj2\pi f_{hm} t},~0\le t-h{T} \le T.
	\end{align}
	{Throughout the paper, we consider that the radar transmitter is equipped with a uniform linear array with the antenna spacing of half a wavelength.}
	{Refer to Appendix \ref{app: radar signal processing} for an elaboration on FH-MIMO radar signal processing.}

\subsection{Information Embedding for Communications}\label{subsec: information embedding}

Let $ \phi $ denote the angle-of-departure (AoD) of the LoS path with respect to (w.r.t.) the radar transmitter array. The multiple-input-single-output channel response vector is $ \beta \mathbf{a}(\phi) $, where $ \beta $ is the LoS path gain and $ \mathbf{a}(\phi) $ is the steering vector in the direction of $ \phi $. 
(By assuming the pseudo-static channel, we have suppressed the dependence of channel parameters, i.e., $ \beta $ and $ \phi $, on time.) Assuming perfect timing, the signal received by the communication receiver at hop $ h $ is \cite{DFRC_FHcodeSel2018}
\begin{align}\label{eq: comm-received signal continuous}
& y_h(t) = \beta \sum_{m=0}^{M-1}e^{-\mj {\pi m\sin\phi}} {F_{hm}}e^{-\mj2\pi f_{hm} t}  + \xi(t),
\end{align}
where $ F_{hm} $ denotes an information modulation term multiplied to the radar signal at hop $ h $ and antenna $ m $ and $ \xi(t) $ is an additive white Gaussian noise (AWGN).

In PSK, $ F_{hm}(t)= e^{\mj{\varpi}_{hm}}~\forall t $ is taken at hop $ h $ and antenna $ m $, where $ {\varpi}_{hm}\in \Omega_J $ $ (J\ge 1) $ with $ \Omega_J=\left\{  0,\frac{2\pi}{2^J},\cdots,\frac{2\pi(2^J-1)}{2^J} \right\} $ denoting the $ J $-bit PSK constellation.
Besides a perfect synchronization, PSK decoding also requires \cite{DFRC_AmbiguityFunc2018Amin,FH_MIMO_Radar2019_RadarConf}:
	1) {{the pairing between hopping frequencies and radar transmitter antennas}; and}	
	2) {$ \beta $ and $ \phi $}.	
	FHCS exploits the different combinations of hopping frequencies to convey information bits \cite{DFRC_FHcodeSel2018}. 
	Since $ M(< K) $ out of $ K $ frequencies are selected per hop, there are $ C_K^M $ different combinations of \HFs. These combinations are used as constellation symbols. 	
	At a communication receiver, 
	only a Fourier transform is performed on the received signal to identify the \HFs~and then demodulate an FHCS symbol \cite{DFRC_FHcodeSel2018}. 
However, to ensure the waveform orthogonality given in (\ref{eq: fphm ne fphm'}), a perfect synchronization is required in \cite{DFRC_FHcodeSel2018}.

\subsection{Proposed Frame Structure}\label{subsec: proposed frame structure}

We consider that radar varies hopping frequencies without notifying the communication receiver. To fulfill data communications and reduce overhead, we assume packet communications and propose the signal frame structure as shown in Fig. \ref{fig: signal structure}(b). In each frame, the first two hops are set identical to enable effective estimation of timing offset, carrier frequency offset (CFO) and channel; and the remaining hops are used for data transmission. 

\begin{figure}[!t]
	\centerline{\includegraphics[width=80mm]{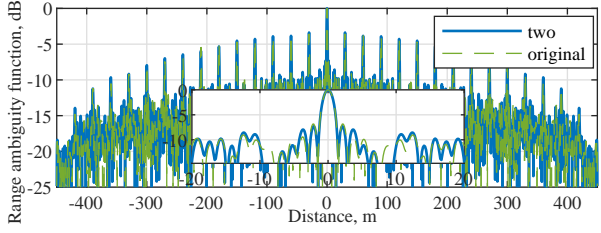}}%
	\caption{The impact of using two identical hops on the range ambiguity function (RAF) of an FH-MIMO radar, where, with reference to \cite{DFRC_FHcodeSel2018}, the radar is configured as $ M=10 $, $ K=20 $, $ H=15 $, $ T=0.2~\mu $s, $ f_{\mathrm{L}}=8 $ GHz and $ B=100 $ MHz. For the original RAF, the \HFs, $ f_{hm}~\forall h,m $, are randomly selected from $ \{f_k(=kB/K)~\forall k\} $.}
	\label{fig: ambiguity function with insertion of best HFs}
	\vspace{-10pt}
\end{figure}

Using two identical hops at a radar pulse can affect the range ambiguity function but only slightly. Fig. \ref{fig: ambiguity function with insertion of best HFs} compares the range ambiguity function of an FH-MIMO radar having two identical hops at the beginning or not, where the sub-optimal \HF~sequence, denoted by the $ M\times 1 $ vector $ \mathbf{f}^* $, are used. ($ \mathbf{f}^* $ will be designed in Section \ref{subsec: design sub-optimal HFs}). 
We see from Fig. \ref{fig: ambiguity function with insertion of best HFs} that only slight changes are incurred to the sidelobes of the range ambiguity function with the mainlobe unaffected by using $ \mathbf{f}^* $ at the the first two hops. 

The use of two identical hops enables simple and effective coarse timing and CFO estimation by performing the conventional energy-based or autocorrelation packet detection \cite{DFRC_ChnnlEst_FanLiu_howMuchInfo}. A coarse timing offset, denoted by $ \eta(>0) $, can be readily estimated at each frame. When the CFO is not very large, a high-accuracy CFO estimation can be readily achieved based on the well-developed methods in the literature; refer to \cite[Chapter 5.4]{book_mimoOFDMmatlab} for a review. {(Details of the coarse timing and CFO estimation are out of the scope of this paper.)} To this end, we take a zero CFO and a non-zero timing offset of $ \eta(>0) $.

\begin{figure*}[!t]\small
	\begin{align}\label{eq: communication signal p h digitized}%
	& y_{h}(i) =  \beta \times \left\{ 
	\begin{array}{ll}
	\sum_{m=0}^{M-1}F_{hm} {e^{-\mj{\pi m\sin\phi}}}
	e^{-\mj2\pi \frac{k_{hm}B}{K} (iT_\mathrm{s}+{\eta}) }+\xi(i),&\text{for  }i= 0,1,\cdots,{{L}}-L_{\eta}-1,\\
	\sum_{m=0}^{M-1}F_{(h+1)m} {e^{-\mj{\pi m\sin\phi}}}e^{-\mj2\pi \frac{k_{(h+1)m}B}{K} ((i-{{L}})T_\mathrm{s} + {\eta}) }+\xi(i),&\text{for }i={{L}}-L_{\eta},\cdots,L-1,
	\end{array}
	\right.  
	\end{align}
	\vspace{-10pt}
\end{figure*}

\subsection{Problem Formulation}\label{subsec: problem formulation}

{At the communication receiver, the RF signal is down converted to the baseband using a local oscillator signal with the frequency $ f_{\mathrm{L}} $. Then, the signal is sampled at a sampling interval of $ T_{\mathrm{s}} $.} Each hop has $ L(=\frac{T}{T_{\mathrm{s}}}) $ samples.
Affected by the timing offset $ \eta $, the initial sampling point of each hop is delayed by $ L_{\eta}(=\lfloor\frac{\eta}{T_s}\rceil) $ samples, where $ \lfloor x\rceil $ rounds $ x $ to the nearest integers. Based on (\ref{eq: comm-received signal continuous}), 
the $ i $-th $ (i=0,1,\cdots,L-1) $ digitized communication signal at hop $ h $, denoted by $ y_h(i) $, is given by (\ref{eq: communication signal p h digitized}), where $ (f_{hm}-f_{\mathrm{L}}) $ is replaced with $ \frac{k_{hm}B}{K} $, $ k_{hm}(\in\{0,1,\cdots,K-1\}) $ denotes the selected sub-band for antenna $ m $ at hop $ h $, and $ (f_{(h+1)m}-f_{\mathrm{L}}) $ is replaced similarly.

{We see from (\ref{eq: communication signal p h digitized}) that extracting $ F_{hm} $ for communication decoding can be non-trivial due to the disturbing phases caused by $ k_{hm} $, $ \eta $, $ \phi $ and $ \beta $.} Although the $ M $ \HFs~at hop $ h $ can be estimated from the discrete Fourier transform (DFT) of $ y_h(i) $, determining $ k_{hm} $ requires the pairing between the \HFs~and antennas.  Acquiring the pairing information and updating the information~as frequently as the primary radar does (i.e., $ H $ times per radar PRI) can be challenging.

We also see from (\ref{eq: communication signal p h digitized}) that the phases incurred by $ \phi $ and $ \eta $ are coupled in a multiplicative manner for each antenna $ m $.
This is drastically different from the conventional communications with a constant $ \eta $ across antennas and hence invalidates conventional methods for estimating $ \eta $ and $ \phi $, e.g., in \cite{book_mimoOFDMmatlab}.
The coupling destroys the linear phase relation in $ e^{-\mj{\pi m\sin\phi}} $ (across $ m $), as required for $ \phi $ estimation \cite{MUSIC_godara1997application}.  
On the other hand, due to the random, independent frequency hopping across hops, the exponential term $ e^{-\mj2\pi \frac{k_{hm}B\eta}{K} } $ is also random across hops.
These factors make the joint estimation of $ \eta $, $ \phi $ and $ \beta $ challenging.

\section{Novel FH-MIMO Waveform for Channel Estimation in DFRC}
\label{sec: new waveform}
In this section, we first design a novel FH-MIMO radar waveform and then depict the proposed channel estimation scheme in overall.

\subsection{Novel FH-MIMO Radar Waveform} \label{subsec: novel waveform and HF estimation}
{Due to the waveform orthogonality given in (\ref{eq: fphm ne fphm'}), the $ M $ signals transmitted from the $ M $ radar antennas have different centroid frequencies. This indicates that the $ M $ signals can be differentiated in the frequency domain.}
{Thanks to the two identical hops at the beginning of each PRI (see Fig. \ref{fig: signal structure}(b)), the waveform orthogonality condition given in (\ref{eq: fphm ne fphm'}) can be ensured, as analyzed below.}
Taking the $ L $-point DFT of the digitized samples of hop $ h $, i.e., $ y_h(i)~(i=0,1,\cdots,L-1) $ given in (\ref{eq: communication signal p h digitized}), the frequency-domain received signal at the $ l $-th discrete frequency $ \frac{l}{LT_{\mathrm{{s}}}} $, denoted by $ Y_{h}(l) $, is 
\begin{align}\label{eq: DFT of digitized communication signal}
&Y_{h}(l) = \sum_{i=0}^{L-1} y_{h}(i)e^{-\mj\frac{2\pi il}{L}}=L\beta  \sum_{m=0}^{M-1} {e^{-\mj{\pi m\sin\phi}}} e^{-\mj2\pi \frac{k_{hm}B\eta}{K} }\times \nonumber\\
&  e^{-\mj\frac{\pi(L-1)l }{L}} \delta\left( l- \Big( L-{k_{hm}BT}/{K} \Big) \right)e^{-\mj\frac{\pi(L-1) k_{hm}BT}{KL}} + \Xi(l),
\end{align}
where $ \delta(l) $ denotes the Dirac delta function and $ \Xi(l) $ is the DFT of the AWGN $ \xi(i) $. 
{Note that the summation term in (\ref{eq: DFT of digitized communication signal}), as indexed by $ m $, is solely related to the signal transmitted by radar antenna $ m $. These terms will be separated and used for the estimation methods to be developed in Sections \ref{sec: tau estimaiton} and \ref{sec: estimate u and beta and perform communciations}.}

\begin{figure}[!t]
	\centerline{\includegraphics[width=80mm]{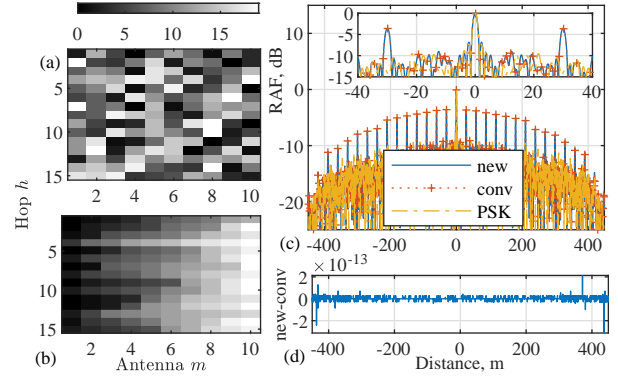}}%
	\caption{(a) The \HFs~of a conventional FH-MIMO radar waveform given in (\ref{eq: transmitted signal p h m}); (b) The \HFs~of the new waveform given by (\ref{eq: new waveform}); (c) Comparison of the range ambiguity function (RAF) of the FH-MIMO radar using the conventional, new waveforms and PSK-embedded new waveform; and (d) Difference between the first two RAFs from Fig. \ref{fig: range ambiguity function}(c). 
		The same radar configuration as in Fig. \ref{fig: ambiguity function with insertion of best HFs} is used here.}
	\label{fig: range ambiguity function}
\end{figure}

Due to the delta function in (\ref{eq: DFT of digitized communication signal}), $ M $ peaks of $ |Y_h(l)| $ can be detected at $ L-{k_{hm}BT}/{K}~(m=0,1,\cdots,M-1) $. By identifying the $ M $ largest peaks of $ |Y_h(l)| $, the set of $ \{k_{hm}~\forall m\} $ can be obtained. However, we cannot determine the pairing between the \HFs~and the radar transmitter antennas, as $ k_{hm}~\forall m $ can take $ \forall k(\in[0,K-1]) $ in conventional FH-MIMO radars \cite{DFRC_AmbiguityFunc2018Amin,Amb_FH_MIMO2008TSP}. 
To solve this problem, we design a novel waveform by introducing a re-ordering of \HFs~at any hop in an ascending order\footnote{It can also be a descending order, which does not affect the property of the new waveform to be unveiled in Proposition \ref{pp: range ambiguity} and the estimation methods to be proposed in Sections \ref{sec: tau estimaiton} and \ref{subsec: estimation of u and beta}.}, as given by
{\begin{subequations}\label{eq: new waveform}
	\begin{align}
	&\tilde{{s}}_{hm}(t) = e^{\mj2\pi ({f}_{\mathrm{L}}+\tilde{k}_{hm}B/K) t}~\forall h,~0\le t-h{T}\le T\nonumber\\
	\mathrm{s.t.}~~& \tilde{k}_{h0}<\tilde{k}_{h1}<\cdots<\tilde{k}_{h(M-1)},\label{eq: new waveform a}\\
	& \left\{\tilde{k}_{hm}~\forall m\right\} = \left\{{k}_{hm}~\forall m\right\}. \label{eq: new waveform b}
	\end{align} 
\end{subequations}
where both $ \tilde{k}_{hm} $ and $ {k}_{hm} $ denote the sub-band index of the radar-transmitted signal from antenna $ m $ at hop $ h $, and the former is for the new waveform while the later is for the conventional waveform.} 
An illustration of the new waveform is provided in Fig. \ref{fig: range ambiguity function}(b), where the \HFs~across antennas and hops are displayed in scaled gray colors, and the \HFs~of a conventional FH-MIMO waveform are given in Fig. \ref{fig: range ambiguity function}(a) for reference.

Enabled by the new waveform, we can now determine the pairing between the \HFs~and radar transmitter antennas. 
Let $ l_m^* $ denote the index of the $ m $-th peak of $ |Y_{h}(l)| $, satisfying
\begin{align}\label{eq: l0*<l1*<...}
0\le l_0^*<l_1^*<\cdots<l_{M-1}^*\le L-1.
\end{align}
From the parameter of the delta function in (\ref{eq: DFT of digitized communication signal}), we see that a smaller $ k_{hm} $ corresponds to a larger index of the peak. Based on this observation, (\ref{eq: new waveform a}) and (\ref{eq: l0*<l1*<...}), we can estimate $ \tilde{k}_{hm} $ as
\begin{align}\label{eq: k_ph^m}
\hat{k}_{hm} = { \left(\frac{ L-l_{M-1-m}^*}{LT_{\mathrm{\textrm{s}}}}\right) }\Big/{ \left(\frac{B}{K }\right) }=\frac{K (L-l_{M-1-m}^*)}{BT}.
\end{align}
The above estimation is achieved without degrading radar ranging due to the following property.

\vspace{2pt}
\begin{proposition}\label{pp: range ambiguity}
	 \textit{The novel FH-MIMO radar waveform, $ \tilde{{s}}_{hm}(t) $, has the same range ambiguity function as the original FH-MIMO radar based on $ {{s}}_{hm}(t) $ given in (\ref{eq: transmitted signal p h m}).}
\end{proposition}  
\vspace{2pt}

Refer to Appendix \ref{app: proof on invariable range ambiguity function} for the proof of Proposition \ref{pp: range ambiguity}. Fig. \ref{fig: range ambiguity function}(c) compares the range ambiguity functions of an FH-MIMO radar, where the conventional range ambiguity function is calculated by substituting the \HFs~shown in Fig. \ref{fig: range ambiguity function}(a) into (\ref{eq: chi(tau)}) {(given in Appendix \ref{app: proof on invariable range ambiguity function})}, and the new range ambiguity function is calculated based on the \HFs~shown in \ref{fig: range ambiguity function}(b). 
We see from Fig. \ref{fig: range ambiguity function}(c) that the new range ambiguity function overlaps with the conventional one. This is further validated by Fig. \ref{fig: range ambiguity function}(d). {Fig. \ref{fig: range ambiguity function}(c) also plots the range ambiguity function using the PSK-embedded new waveform, where $ (M\times H) $ number of randomly generated BPSK symbols are multiplied onto $ \tilde{s}_{hm}~\forall h,m $, one for each. Same as the conventional FH-MIMO waveform \cite{DFRC_AmbiguityFunc2018Amin}, the new waveform, when combined with PSK, can have range sidelobe spikes suppressed.
	This is because incoherent PSK phases can prevent periodic energy accumulations in range sidelobes; refer to  \cite{DFRC_AmbiguityFunc2018Amin} for an in-depth analysis of the spike suppression.
}

\subsection{Overall Channel Estimation Scheme}\label{subsec: overall channel estimation scheme}
Using the new FH-MIMO radar waveform, we develop a channel estimation scheme by first focusing on LoS channels and a single-antenna communication receiver. Then, the proposed scheme is extended to multi-path channels in Section \ref{sec: extensions to NLOS and multi-antenna scenarios}.
The scheme includes two steps, which will be detailed in Sections \ref{sec: tau estimaiton} and \ref{subsec: estimation of u and beta}, respectively. 
{The estimation methods to be proposed can be performed based on a single hop, i.e., the sampled hop $ 0 $ as highlighted in Fig. \ref{fig: signal structure}.} Hence, we drop the subscript ``$ (\cdot)_h $'', unless otherwise specified.

\subsubsection{Estimate $ \omega_{{\eta}} $, a function of the timing offset $ \eta $}\label{subsubsec: tau estimation}
{Substituting (\ref{eq: l0*<l1*<...}) and (\ref{eq: k_ph^m}) into (\ref{eq: DFT of digitized communication signal}), the signal from the $ m $-th radar transmitter antenna can be extracted, as given by 
	\begin{align} \label{eq: DFT of com signal m-th frequency}
	Y_m  = Y(l_{M-1-m}^*)   
	= \tilde{\beta}  e^{-\mj\frac{2\pi m u}{M}}   \omega_{\eta}^{\hat{k}_{m}} + \Xi(l_{M-1-m}^*)
	\end{align}
	where the intermediate variables, $ \tilde{\beta} $, $ u $ and $ \omega_{\eta} $, are defined as
	\begin{align}\label{eq: intermediate variables}
	\tilde{\beta} \triangleq L{{\beta}  e^{-\mj{\pi(L-1)}}},~u\triangleq {M\sin\phi}/{2},~\omega_{\eta} \triangleq e^{-\mj2\pi \frac{B\eta}{K}  }.
	\end{align} 
	The way we define $ u $ is to enable accurate estimation of $ \phi $, as will be designed in Section \ref{subsec: estimation of u and beta}.} {Note that $ Y(l_{M-1-m}^*) $ takes the DFT output at the $ l_{M-1-m}^* $-th discrete frequency, while the DFT output given in (\ref{eq: DFT of digitized communication signal}) is calculated based on the $ L $ signal samples in the first single hop received by the communication receiver.}

From (\ref{eq: DFT of com signal m-th frequency}), we see that $ \omega_{\eta}^{\hat{k}_{m}} $ $ \forall  m $ is multiplied to $ e^{-\mj\frac{2\pi m u}{M}} $ $ \forall  m $ pointwise. 
Due to the phase disturbance caused by $ \omega_{\eta}^{\hat{k}_{m}} $, the linear phase relation in $ e^{-\mj\frac{2\pi m u}{M}} $, which is the key for angle estimation \cite{MUSIC_godara1997application}, is scrambled. 
Nevertheless, we notice that the linear phase relation in $ e^{-\mj\frac{2\pi m u}{M}} $ can be exploited to suppress the impact of $ u $ on $ \eta $ estimation. This enables us to estimate $ \omega_{\eta} $ unambiguously, as to be designed in Section \ref{sec: tau estimaiton}. We propose to estimate $ \omega_{\eta} $ rather than $ \eta $ due to the non-trivial phase ambiguity issue in $ \eta $ estimation. This will be clear in Section \ref{sec: tau estimaiton}. It is noteworthy that using the estimate of $ \omega_{\eta} $ is sufficient to suppress the impact of $ \eta $ on $ \phi $ estimation and communication decoding.

\subsubsection{Estimate channel parameters $ \phi $ and $ \beta $}\label{subsubsec: u and beta estimation}
Given the estimate of $ \omega_{\eta} $, we can remove $ \omega_{\eta}^{\hat{k}_{m}}~\forall m $ in (\ref{eq: DFT of com signal m-th frequency}) to further estimate $ \phi $ and $ \beta $. It is non-trivial to estimate $ \phi $ based on the single-hop signals $ Y_m~\forall m $. 
The spatial searching methods developed in \cite{Kai_UCA2019Qiang_CL,FFG_OneSnapshot2017CL} can be performed using a single-snapshot. However, these methods \cite{Kai_UCA2019Qiang_CL,FFG_OneSnapshot2017CL} 
can be time-consuming in achieving a satisfactory estimation accuracy, since more searching grids are required for a better angle resolution. 
In a different yet related context, we developed in \cite{Kai_freqEst2020CL} a method to estimate the frequency of a single-tone exponential signal having the similar expression to the $ u $-related term in (\ref{eq: DFT of com signal m-th frequency}). The method is low in computational complexity and efficient in the sense of approaching the Cram\'er-Rao lower bound (CRLB). 
{Thus, we apply the method \cite{Kai_freqEst2020CL} to develop an accurate estimation method for $ \phi $ and $ \beta $, as will be elaborated on in Section \ref{subsec: estimation of u and beta}.  }

\section{Proposed Method for $ \omega_{\eta} $ Estimation}\label{sec: tau estimaiton}

As illustrated in Section \ref{subsec: overall channel estimation scheme}, the estimation of $ \omega_{\eta} $ is key to the overall channel estimation. In this section, we first develop two estimators for $ \omega_{\eta} $, 
	then derive their MSELBs, and moreover design a sub-optimal \HF~sequence.

\subsection{Estimation of $ \omega_{\eta} $}\label{subsec: estimation of eta}
To estimate $ \omega_{\eta} $ based on $ {Y}_m $ given in (\ref{eq: DFT of com signal m-th frequency}), we need to suppress the impact of $ \tilde{\beta}  $ and $  u  $.
Since $ \tilde{\beta} $ is independent of $ m $, we can suppress $ \tilde{\beta} $ by taking the ratio of adjacent $ {Y}_m $, 
\begin{align}\label{eq: breve{Y}}
	& \breve{Y}_m = \frac{{Y}_m}{{Y}_{m+1} } = e^{\mj\frac{2\pi u}{M}}\omega_{\eta}^{\hat{k}_{m}-\hat{k}_{(m+1)}},~m=0,1,\cdots,M-2.
\end{align}
We see from (\ref{eq: breve{Y}}) that the $ u $-related term is now independent of $ m $, and hence, by taking the ratio of adjacent $ \breve{Y}_m $, the impact of $ u $ can be suppressed, i.e., 
\begin{align}\label{eq: barY}
&\bar{Y}_m = \frac{\breve{Y}_m}{\breve{Y}_{m+1} } =  \omega_{\eta}^{\kappa_{m}}=e^{\mj\kappa_{m}  \angle\omega_{\eta}},~m=0,\cdots,M-3\nonumber\\
&\mathrm{s.t.}~\kappa_{m}\triangleq\hat{k}_{m}-2\hat{k}_{(m+1)}+\hat{k}_{(m+2)},
\end{align}
where $ \angle\omega_{\eta} $ takes the phase of $ \omega_{\eta} $ in the interval of $ [-\pi,\pi] $. By estimating $ \angle\omega_{\eta} $, $ \omega_{\eta} $ can be determined. 

We see from (\ref{eq: barY}) that $ \angle\omega_{\eta} $ can be estimated by taking the phase of $ \bar{Y}_m $, which, however, requires $ \kappa_{m}\ne 0 $. Based on (\ref{eq: k_ph^m}), we can identify the set of antenna indexes satisfying 
\begin{align}\label{eq: mathcal M}
	\mathcal{M}=\{\forall \tilde{m}\},~~\mathrm{s.t.}~\kappa_{\tilde{m}}\ne 0 .
\end{align}
We also see from (\ref{eq: barY}) that directly taking the angle of $ \bar{Y}_m $ can lead to phase ambiguity due to potential cases of $ |\kappa_{m}\angle\omega_{\eta}|>\pi $. Hence, we have two possible estimates of $ \angle\omega_{\eta} $, i.e.,
\begin{align}\label{eq: hat eta m}
	{\measuredangle\omega}_{\eta}(d_{\tilde{m}}) = \left\{ 
	\begin{array}{cl}
		\kappa_{\tilde{m}} \measuredangle \bar{Y}_{\tilde{m}}~(d_{\tilde{m}}=0),&\mathrm{if}~|\kappa_{\tilde{m}}| = 1\\
		\frac{\measuredangle \bar{Y}_{\tilde{m}} + 2d_{\tilde{m}} \pi}{\kappa_{\tilde{m}}}~(d_{\tilde{m}}=0, \pm1,\cdots),& \mathrm{otherwise}
	\end{array}
	\right.
\end{align}
where $ d_{\tilde{m}} $ is the ambiguity degree.

From (\ref{eq: hat eta m}), we see that, if $ |\kappa_{\tilde{m}}| = 1 $ holds for some $ \tilde{m} $, we obtain the unambiguous estimate of $ \angle\omega_{\eta} $ directly. 
Let $ \bar{\mathcal{M}}\subseteq \mathcal{M} $ denote the set of $ \bar{m} $ such that $ |\kappa_{\bar{m}}|=1 $, i.e., 
\begin{align}\label{eq: bar mathcal M}
	\bar{\mathcal{M}} = \{\forall\bar{m}\}\subseteq\mathcal{M}~~\mathrm{s.t.}~{|\kappa_{\bar{m}}|=1~\forall \bar{m}\in{\mathcal{M}}}.
\end{align}
Substituting (\ref{eq: bar mathcal M}) into (\ref{eq: hat eta m}), we can accumulate $ \bar{Y}_{\bar{m}} $ coherently
across $ \bar{m} $ and then take the angle for $ \angle\omega_{\eta} $ estimation. This leads to the first estimator of $ \angle\omega_{\eta} $, referred to as the {coherent accumulation estimator (CAE)}:
\begin{align}\label{eq: eta estimate based on bar mathcal M}
	\text{CAE: }{\measuredangle}\bar{\omega}_{\eta} = \angle \Big(\frac{1}{\bar{M}}\sum_{\bar{m}\in \bar{\mathcal{M}} }  \left(\Re\{\bar{Y}_{\bar{m}}\}+\mj\kappa_{\bar{m}}\Im\{\bar{Y}_{\bar{m}} \}\right)\Big),
\end{align}
where $ \bar{M} $ is the dimension of the set $ \bar{\mathcal{M}} $. Note that $ \kappa_{\bar{m}} $ is multiplied to $ \Im\{\bar{Y}_{\bar{m}}\} $, as $ \kappa_{\bar{m}}=-1 $ can happen for some $ \bar{m} $. 

Depending on the \HFs, we can have $ \bar{\mathcal{M}}=\emptyset $, which clearly invalidates CAE given in (\ref{eq: eta estimate based on bar mathcal M}). 
In this case, we can still estimate $ \angle\omega_{\eta} $ by removing the estimation ambiguity in (\ref{eq: hat eta m}). Specifically, we can exploit the Chinese remainder theorem \cite{CRT_TVT2011} to suppress the ambiguity in (\ref{eq: hat eta m}). To do this, we need to identify the set $ \breve{\mathcal{M}} $ rendering $ \left\{|\kappa_{\breve{m}}|~\forall \breve{m} \in \breve{\mathcal{M}} \right\} $ co-prime with at least two elements, i.e., 
\begin{align} \label{eq: breve mathcal M}
	\breve{\mathcal{M}}=\{\forall \breve{m}\}~~\mathrm{s.t.}~\mathcal{G}\left\{|\kappa_{\breve{m}}|(\ne1)~\forall \breve{m}\in \mathcal{M}\right\}=1,~\breve{M}\ge 2,
\end{align}
where $ \mathcal{G}\{\cdot\} $ takes GCD, and $ \breve{M} $ denotes the dimension of the set $ \breve{\mathcal{M}} $. 
By identifying the ambiguity degree $ d_{\breve{m}}^*~\forall \breve{m} $ such that the estimates $ {\measuredangle}\omega_{\eta}(d_{\breve{m}}^*)~\forall \breve{m}\in \breve{\mathcal{M}} $ are identical, the second estimator for $ \angle\omega_{\eta} $, referred to as the {Chinese remainder theorem estimator (CRE)}, is achieved:
\begin{align} \label{eq: eta estimate based on breve mathcal M}
\text{CRE: }{\measuredangle}\breve{\omega}_{\eta} = \frac{1}{\breve{M}}\sum_{\breve{m}\in \breve{\mathcal{M}} } {\measuredangle}\omega_{\eta}(d_{\breve{m}}^*).
\end{align}
Note that CAE and CRE have their own favorable working conditions and correspondingly different estimation accuracy, as analyzed below.

\subsection{Performance Analysis and Comparison of the Estimators}
To compare the two estimators, we first derive the MSELBs of them. 
In the following, the SNR, denoted by $ \gamma $, refers to the ratio between the received signal power and the communication receiver noise power. Based on (\ref{eq: communication signal p h digitized}), we have $ \gamma=|\beta|^2/\sigma_n^2 $, where $ \sigma_n^2 = {\mathbb{E}\{|\xi(i)|^2\}} $ is the noise variance of the AWGN $ \xi(i) $. The high-SNR MSELBs of CAE and CRE are derived as follows\footnote{{Note that CRLB is the lower limit of the MSELB derived here; while, according to \cite{CRLB_MLEphaseEstimator1997TCOM}, CRLB is not applicable to estimators, like the proposed CAE and CRE, which estimate a random phase with a finite support $ [-\pi,\pi) $.} }.

\vspace{2pt}
\begin{proposition}\label{pp: mselbs}
	\textit{At high SNR, the MSELBs of CAE and CRE are (\ref{eq: mselb of bar mathcal M}) and (\ref{eq: mselb of breve mathcal M}), respectively.
\begin{gather}
	\bar{\sigma}^2_{\eta} = {3}/({\bar{M}L\gamma})\label{eq: mselb of bar mathcal M}\\
	\breve{\sigma}^2_{\eta} = \frac{1}{\breve{M}^2} \sum_{\breve{m}\in\breve{\mathcal{M}}} {3}/({\kappa_{\breve{m}}^2 L\gamma})
	\label{eq: mselb of breve mathcal M}
\end{gather}
}	
\end{proposition}

Refer to Appendix \ref{app: proof of MSELBs} for the proof of Proposition \ref{pp: mselbs}.
We see from (\ref{eq: mselb of bar mathcal M}) and (\ref{eq: mselb of breve mathcal M}) that the accuracy of both estimators are dependent on hopping frequencies. Specifically, $ \bar{\sigma}^2_{\eta} $ decreases when the number of ones in $ \bar{\mathcal{M}} $ (i.e., $ \bar{M} $) increases. Thus, the MSELB for CAE has a lower limit, i.e., $ \underline{\bar{\sigma}^2_{\eta}} = \frac{3}{(M-2)L\gamma}\le \bar{\sigma}^2_{\eta},  $
where $ (M-2) $ is the maximum value that $ \bar{M} $ can take. 
In contrast, the accuracy of CRE depends on the number and the values of the co-prime elements
in $ \breve{\mathcal{M}} $.
Given (\ref{eq: breve mathcal M}), $ \breve{{\mathcal{M}}}=\{2,3\} $ with $ \breve{M}=2 $ is the smallest set with the minimum co-prime numbers. Substituting $ \breve{{\mathcal{M}}}=\{2,3\} $ into (\ref{eq: mselb of breve mathcal M}), we obtain the upper limit of $ \breve{\sigma}^2_{\eta} $, as given by $ \overline{\breve{\sigma}^2_{\eta}} =\frac{2}{L\gamma}\times  \frac{\frac{1}{4}+\frac{1}{9}}{4} =\frac{3}{L\gamma}\times  \frac{13}{144} \ge \breve{\sigma}^2_{\eta} . $ Note that if $ \underline{\bar{\sigma}^2_{\eta}}>\overline{\breve{\sigma}^2_{\eta}} $ then $ \bar{\sigma}^2_{\eta}>\breve{\sigma}^2_{\eta} $ is assured. Moreover, $ \underline{\bar{\sigma}^2_{\eta}}>\overline{\breve{\sigma}^2_{\eta}} $ leads to $ \frac{1}{M-2}>\frac{13}{144} $, and further $ M\le 13 $. Thus, we have the following corollary,

\vspace{2pt}
\begin{corollary}\label{col: comparison of two estimates}
	\textit{{For a uniform linear array with the antenna spacing of half a wavelength}, provided the number of antennas at the radar transmitter satisfies $ M\le 13 $, CRE always has a better asymptotic performance than CAE, i.e., $ \bar{\sigma}^2_{\eta}>\breve{\sigma}^2_{\eta} $.
	}   
\end{corollary}
\vspace{2pt}

\begin{remark}\label{rmk: low snr comparison of estimates}
	Corollary \ref{col: comparison of two estimates} compares the asymptotic performance of the two estimates in high SNR regions, where the ambiguity degree $ d_{\breve{m}}^* $ required for CRE can be reliably identified; see (\ref{eq: eta estimate based on breve mathcal M}). 
	In low SNR regions, however, the correct identification of $ d_{\breve{m}}^* $ cannot be ensured, which degrades the estimation accuracy of CRE. This issue does not exist for CAE which does not have estimation ambiguity. Moreover, when $ \bar{M} $ is large, the coherent accumulation in (\ref{eq: eta estimate based on bar mathcal M}) can help improve the estimation SNR of CAE. In this sense, CAE is more suited for low SNR regions, compared with CRE. 
	As to be observed from simulation in Section \ref{sec: simulations}, there is an SNR threshold of $ \gamma $, denoted by $ \gamma_{\mathrm{T}} $, satisfying:
	if $ \gamma>\gamma_{\mathrm{T}} $, CRE is more accurate than CAE; otherwise, CAE is better.  
\end{remark}

\subsection{Design of a Sub-optimal Hopping Frequency Sequence}\label{subsec: design sub-optimal HFs}

{At the communication receiver, $ \gamma $ can be estimated. By comparing $ \gamma $ and $ \gamma_{\mathrm{T}} $, the receiver can choose which estimator to use between CAE and CRE.}
However, this does not apply to the radar transmitter with no \textit{a-priori} information on $ \gamma $. 
{To this end, it is necessary to design a \HF~sequence which renders both estimators applicable at the communication receiver.}
Such a sequence is optimal when the MSELB of CAE, $ \bar{\sigma}^2_{\eta} $, and that of CRE, $\breve{\sigma}^2_{\eta} $, are minimized simultaneously. {The optimality, however, cannot be achieved}, since $ \bar{M}=(M-2) $ is required to minimize $ \bar{\sigma}^2_{\eta} $; whereas $ \bar{M} $ can take no greater than $ (M-4) $ to ensure at least two elements in $ \breve{{{\mathcal{M}}}} $ for CRE.
Next, we propose a sub-optimal design of \HF~sequence that ensures the largest coherent accumulation gain of $ (M-4) $ in low SNR regions and accordingly minimizes the MSELB of CRE.
Let $ \mathbf{f}^* $ denote the sub-optimal \HF~sequence to be designed. Since $ \mathbf{f}^* = f_{\mathrm{L}}+B\mathbf{k}^*/K $, where $ \mathbf{k}^* = [k_0^*,k_1^*,\cdots,k_{M-1}^*]^{\mathrm{T}} $, we can design $ \mathbf{k}^* $ equivalently.

\textit{1) Minimizing $ \bar{\sigma}^2_{\eta} $:}
The minimization of the MSELB of CAE, $ \bar{\sigma}^2_{\eta} $, can be achieved at $ \bar{M}=(M-4) $, i.e., having $ (M-4) $ elements in $ \bar{\mathcal{M}} $. 
This requires at least $ (M-2) $ \HFs, since according to (\ref{eq: barY}), one element of $ \bar{\mathcal{M}} $ is calculated using three \HFs. 
{Based on (\ref{eq: barY}), we design the following recursive calculation of $ k_m^* $ to ensure $ \bar{M}=M-4 $, }
\begin{subequations}\label{eq: fm+2=2fm+1-fm...}
	\begin{align}
		&~k_{m+2}^* = 2k_{m+1}^*-k_m^*\pm 1\label{eq: fm+2=2fm+1-fm...a}\\
		\mathrm{s.t.}~~&~k_{m+2}^*>k_{m+1}^*,~m=0,1,\cdots,M-5\label{eq: fm+2=2fm+1-fm...b}\\
		&~k_0^*=0,~k_1^*=1,	\label{eq: fm+2=2fm+1-fm...c}
	\end{align}
\end{subequations}
where the constraint (\ref{eq: fm+2=2fm+1-fm...b}) complies with the constraint (\ref{eq: new waveform a}) of the new FH-MIMO radar waveform; and (\ref{eq: fm+2=2fm+1-fm...c}) initializes the first two \HFs~associated with antennas $ m=0 $ and $ 1 $. 
Given the recursive calculation in (\ref{eq: fm+2=2fm+1-fm...a}), taking the minimum values for $ k_0^* $ and $ k_1^* $ also minimizes $ k_{M-3}^* $. The minimization of $ k_{M-3}^* $ is important for the design of the remaining two elements to minimize the MSELB of CRE, $ \breve{\sigma}^2_{\eta} $, as elaborated on below.

\textit{2) Minimizing $ \breve{\sigma}^2_{\eta} $:}
By solving (\ref{eq: fm+2=2fm+1-fm...}), 
the first $ (M-2) $ elements~in $ \mathbf{k}^* $ are determined, which leaves $ k_{M-2}^* $ and $ k_{M-1}^* $ to be designed for minimizing $ \breve{\sigma}_{\eta}^2 $. 
Moreover, $ k_{M-2}^* $ and $ k_{M-1}^* $ can only be selected from $ {\mathcal{K}}=\{k_{M-3}^*+1,k_{M-3}^*+2\cdots,K-1\} $.
According to (\ref{eq: mselb of breve mathcal M}), the problem of minimizing $ \breve{\sigma}^2_{\eta} $ is turned into: {the selection of two elements from $ {\mathcal{K}} $ as the last two elements of $ \mathbf{k}^* $, so that the last four elements of $ \mathbf{k}^* $ can produce two co-prime numbers to minimize $ \rho = \frac{1}{(\breve{M})^2}  \sum_{\breve{m}\in\breve{\mathcal{M}}} \frac{1}{\kappa_{\breve{m}}^2 } $ and hence $ \breve{\sigma}^2_{\eta} $.}
Let $ \{\mathbf{k}_b,~ b=0,1,\cdots,C_{|\mathcal{K}|}^2-1\} $ denote the set for the combinations of selecting two elements from $ {\mathcal{K}} $, where $ |\mathcal{K}| $ is the cardinality of $ \mathcal{K} $. By substituting $ [k_{M-4}^*,k_{M-3}^*,\mathbf{k}_b^{\mathrm{T}}]^{\mathrm{T}} $ into (\ref{eq: barY}) and (\ref{eq: breve mathcal M}), the obtained set of co-prime numbers is denoted by $ \breve{{{\mathcal{M}}}}_b $. Its dimension is $ \breve{{{M}}}_b $. Thus, $ \breve{\sigma}^2_{\eta} $ can be minimized via solving 
\begin{align}\label{eq: mathbf fb*}
	\{\mathbf{k}_{b^*},\rho_{b^*}\}:~~\min_{b\in\{0,1,\cdots,C_{|\mathcal{K}|}^2-1\}}~ \rho_b = \frac{1}{\breve{M}_b^2}  \sum_{\breve{m}\in\breve{\mathcal{M}}_b} \frac{1}{\kappa_{\breve{m}}^2 }.
\end{align}
{Based on (\ref{eq: fm+2=2fm+1-fm...}) and (\ref{eq: mathbf fb*}), the proposed sub-optimal \HF~sequence~is obtained as $ \mathbf{f}^*=f_{\mathrm{L}}+B\mathbf{k}^*/K $, where }
\begin{align}\label{eq: f* and k*}
	\mathbf{k}^* = \left[\UOLoverbrace{k_0^*,k_1^*,\cdots,}[k_{M-4}^*,k_{M-3}^*,]^{\mathrm{(\ref{eq: fm+2=2fm+1-fm...}):~minimizing}~\bar{\sigma}^2_{\eta}}\UOLunderbrace{~\mathbf{k}_{b^*}^{\mathrm{T}}}_{\mathrm{(\ref{eq: mathbf fb*}):~minimizing}~\breve{\sigma}^2_{\eta}}\right]^{\mathrm{T}}.
\end{align}

\section{Estimation of $ \phi $ and $ \tilde{\beta} $}\label{sec: estimate u and beta and perform communciations}

With $ \omega_{\eta} $ estimated, we proceed to develop the method for estimating $ \phi $ and $ \tilde{\beta} $. 
Then, a complexity analysis is provided for the proposed channel estimation scheme.

\subsection{Estimation of $ \phi $ and $ \tilde{\beta} $}\label{subsec: estimation of u and beta}

Based on the estimation obtained in (\ref{eq: eta estimate based on bar mathcal M}) or (\ref{eq: eta estimate based on breve mathcal M}), we obtain the estimate of $ \omega_{\eta} $ as $ \hat{\omega}_{\eta}=e^{\mj\measuredangle \bar{\omega}_{\eta}}\text{ or }e^{\mj\measuredangle \breve{\omega}_{\eta}} $. 
Dividing both sides of (\ref{eq: DFT of com signal m-th frequency}) by $ \hat{\omega}_{\eta}^{\hat{k}_m} $ leads to 
\begin{align}\label{eq: tilde Z}
	{Z}_m = {{Y}_m}\Big/{\hat{\omega}_{\eta}^{\hat{k}_m}} = \tilde{\beta} e^{-\mj\frac{2\pi m u}{M}},
\end{align}
where $ \tilde{\beta} $ and the $ \phi $-related variable, $ u $, are defined in (\ref{eq: intermediate variables}) and the noise term is dropped to focus on algorithm illustration. Note in (\ref{eq: tilde Z}) that $ \omega_{{\eta}}^{\hat{k}_m} $ is assumed to be fully suppressed so that we can focus on formulating the estimation method for $ u $ and $ \tilde{\beta} $. (The impact of the $ {\omega}_{\eta} $ estimation error on the estimations of $ u $ and $ \tilde{\beta} $ will be illustrated in Section \ref{sec: simulations}.)

We see from (\ref{eq: tilde Z}) that $ u $ can be regarded as a discrete frequency, and hence $ u $ estimation is turned into the frequency estimation of a sinusoidal signal $ {Z}_m~(m=0,1,\cdots,M-1) $. The frequency estimator that we developed recently in \cite{Kai_freqEst2020CL} can be applied here for $ u $ estimation. In overall, the estimator first searches for the DFT peak of the sinusoidal signal $ {Z}_m $ to obtain a coarse estimation of $ u $, and then interpolates the DFT coefficients around the peak to refine the estimation. 
Taking the DFT of $ {Z}_m $ w.r.t. $ m $ leads to $ {z}_{m'} = \sum_{m=0}^{M-1}{Z}_me^{-\mj\frac{2\pi m m'}{M}} $. 
By identifying the peak of $ |z_{m'}| $, a coarse estimation of $ u $ can be obtained as $ \frac{ \tilde{m}}{M} $, where $ \tilde{m} $ is the index of the peak. The true value of $ u $ can be written as $ \frac{ \tilde{m}+\delta}{M} $ with $ \delta(\in[-0.5,0.5]) $ being a fractional frequency residual. By estimating $ \delta $, the coarse $ u $ estimate can be refined, as developed below. 
 
We can estimate $ \delta $ recursively from the interpolated DFT coefficients around $ \tilde{m} $. 
Initially, we set $ \delta=0 $, and calculate the interpolated DFT at the discrete frequency $ \tilde{m}\pm\epsilon+\delta $, where $ \epsilon=\min\{M^{-\frac{1}{3}},0.32\} $ \cite[Eq. (23)]{Kai_freqEst2020CL} is an auxiliary variable of the $ u $ estimation algorithm. It has been proved in \cite{FreqEst2019TCOM_Serbes} that the above value of $ \epsilon $ lead to an efficient estimator in the sense of approaching CRLB.
The interpolated DFT coefficients, denoted by $ z_{\pm} $, can be calculated as $ z_{\pm} = \sum_{m=0}^{M-1} {Z}_m e^{-\mj\frac{2\pi m(\tilde{m}+\delta\pm\epsilon)}{M}}. $
An update of $ \delta $ can be obtained using $ z_{\pm} $, i.e., 
\begin{align}\label{eq: delta update}
\delta = \frac{\epsilon\cos^2(\pi\epsilon)}{1-\pi\epsilon\cot(\pi\epsilon)}\times \Re\{\zeta\}+\delta,
\end{align} 
where $ \delta $ on the right-hand side (RHS)  is the old value and $ \zeta=\frac{z_{+}-z_{-}}{z_{+}+z_{-}} $.
Use the new value of $ \delta $ to update $ z_{\pm} $ which is then used, as above, for $ \delta $ update. By updating $ \delta $ three times in overall, the algorithm can generally converge \cite{FreqEst2019TCOM_Serbes}.
The final estimate of $ u $ is obtained as $ \hat{u} = {(\tilde{m}+\delta)}/{M}. $
Substituting $ \hat{u} $ into (\ref{eq: intermediate variables}) and (\ref{eq: tilde Z}), the $ \phi $ and $ \tilde{\beta} $ estimations are
\begin{align}\label{eq: hat beta}
\hat{\phi} = \arcsin \hat{u}\lambda/(Md) ,~\hat{\tilde{\beta}}=\frac{1}{M}\sum_{m=0}^{M-1} {Z}_m e^{\mj\frac{2\pi m\hat{u}}{M}}.
\end{align}

	\subsection{Complexity Analysis}
	In overall, the proposed channel estimation scheme has a low computational complexity, since no computationally intensive operations (e.g., matrix inversion/decomposition) are required. 
	As a matter of fact, the major computations involved in the proposed scheme are an $ L $-dimensional DFT, an $ M $-dimensional DFT and $ N_{\mathrm{iter}} $ numbers of $ M $-dimensional complex vector operations. The first DFT is used for identifying hopping frequencies and extracting the $ M $ peaks that are used for channel estimation; see Sections \ref{subsec: overall channel estimation scheme} and \ref{subsec: estimation of eta}. The second DFT is used for obtaining the coarse estimation of $ u $; see Section \ref{subsec: estimation of u and beta}. The third vector operation is used for refining $ u $ estimation through $ N_{\mathrm{iter}} $ iterations. Thus, the computational complexity of the proposed scheme is given by
	\begin{align}
	\mathcal{O}\left( L'\log L'  + M'\log M'  + M'N_{\mathrm{iter}}  \right)\overset{L\gg M}{\approx} \mathcal{O}\left( L'\log L'  \right),\nonumber
	\end{align}	
	where the fast Fourier transform is used for calculating the two DFTs; $ L'=2^{\lceil \log_2L \rceil} $ and $ M'=2^{\lceil \log_2M \rceil} $; and the last approximation is established as $ N_{\mathrm{iter}} $ is small. As illustrated in Section \ref{subsec: estimation of u and beta}, $ N_{\mathrm{iter}}=3 $ is generally sufficient for the convergence of the $ u $ estimation method.

\section{Applying Channel Estimations to Data Communication}\label{sec: decoding  using estimated channels}
In this section, we illustrate how to apply the estimations of $ \omega_{\eta} $, $ \phi $ and $ \beta $ to perform data communication. To improve the data rate, we propose to combine PSK and FHCS into a new constellation, referred to as PFHCS.
To perform PFHCS modulation at the radar transmitter, we only need to multiply the modulation term $ F_{hm}(t)=e^{\mj{\varpi}_{hm}} $ onto radar waveform, as with the sole PSK; refer to Section \ref{subsec: information embedding}.
{Due to the use of PSK, PFHCS can also suppress the sidelobe spikes in the range ambiguity function of an FH-MIMO radar \cite{DFRC_AmbiguityFunc2018Amin}, as illustrated in Fig. \ref{fig: range ambiguity function}(c).}
At the communication receiver, the PFHCS demodulation can be performed by first demodulating the FHCS sub-symbol and then the PSK sub-symbol.

To demodulate an FHCS symbol, we need to extract the \HFs, i.e., $ k_{hm} $. The subscript $ (\cdot)_h $ is re-added here to differentiate the $ (H-2) $ hops used for data communication.
Referring to Fig. \ref{fig: signal structure}, we see that the sample shift $ L_{\eta} $ caused by the timing offset $ \eta $ needs to be compensated to recover a complete data hop. The value of $ L_{\eta} $ can be extracted from the estimation of $ \angle\omega_{{\eta}} $ obtained in Section \ref{sec: tau estimaiton}. Let $ \measuredangle\omega_{{\eta}} $ denote the $ \angle\omega_{{\eta}} $ estimation which can be either (\ref{eq: eta estimate based on bar mathcal M}) or (\ref{eq: eta estimate based on breve mathcal M}). Based on the definition of $ \omega_{{\eta}} $ given in (\ref{eq: intermediate variables}), the $ \eta $ estimation can be extracted from $ \measuredangle \omega_{\eta} $ as
\begin{align}%
& \hat{{\eta}}_d = {(K\measuredangle\omega_{{\eta}}+2d\pi)}/{(2\pi B)},~\mathrm{s.t.} ~ 0< \hat{{\eta}}_d <T, \nonumber
\end{align}
where $ d(=0,\pm1,\cdots)$ is the ambiguity degree. Note that the constraint can make the number of $ \eta $ estimates limited. 
Using $ \hat{{\eta}}_d $, the sample shift $ L_{\eta} $ can be estimated as $ \hat{L}_{\eta}^{(d)} = \lfloor \hat{{\eta}}_d/T_{\mathrm{s}} \rceil,~d=0,\pm 1,\cdots. $
Given the waveform orthogonality; see (\ref{eq: fphm ne fphm'}), we can remove the estimation ambiguity in $ \hat{L}_{\eta}^{(d)} $.

Re-construct the $ h $-th $ (h=1,\cdots,H-1) $ sampled hop as 
\begin{align}\label{eq: yhd reconstructed hop signals}
&\mathbf{y}_h^{(d)} = \left[ y_{h-1}(L-\hat{L}_{\eta}^{(d)}),\cdots,y_{h-1}(L-1),\right.\nonumber\\
&~~~~~~~~~~~~~~~~~~~~~~~~~\left.y_h(0),\cdots,y_h(L-\hat{L}_{\eta}^{(d)}-1) \right]^{\mathrm{T}}.
\end{align}
By calculating the $ L $-point DFT of $ \mathbf{y}_h^{(d)} $ and searching for the spectrum peaks as done in (\ref{eq: DFT of digitized communication signal}) and (\ref{eq: DFT of com signal m-th frequency}), the $ m $-th peak is denoted by $ Y_{hm}^{(d)} $. 
Similarly, we can calculate the DFT of $ \mathbf{y}_h^{(d)} $ at the $ l_h^* $-th discrete frequency, leading to $ \tilde{Y}_{hm}^{(d)} $.
Here, $ l_h^* $ is taken such that $ l_h^*\notin \{l_{h(M-1-m)}^*~\forall m\} $, where $ l_{h(M-1-m)}^* $ is defined similar to $ l_{M-1-m}^* $ given in (\ref{eq: l0*<l1*<...}).
Provided that $ \mathbf{y}_h^{(d)} $ is correctly re-constructed at $ d=d^* $, we have $ \sum_{m=0}^{M-1}|Y_{hm}^{(d^*)}|=ML $
and $ \sum_{m=0}^{M-1}|\tilde{Y}_{hm}^{(d^*)}|=0$ 
in the absence of noises. The two equations are due to the waveform orthogonality given in (\ref{eq: fphm ne fphm'}). 
Considering inevitable noises, we can identify $ d^* $ robustly via
\begin{align}
d^*:\max_{\left\{\substack{d=0,\\\pm 1,\cdots}\right\}}  \sum_{h=2}^{H-1}\left(\sum_{m=0}^{M-1} \left| Y_{hm}^{(d)} \right| \left/ \sum_{m=0}^{M-1} \left| \tilde{Y}_{hm}^{(d)} \right| \right)\right..
\end{align}

The PFHCS demodulation is summarized below.

\textit{a) FHCS sub-symbol:} After identifying $ d^* $, substitute the indexes of the $ M $ DFT peaks into (\ref{eq: k_ph^m}), producing $ \hat{k}_{hm} $ and the estimate of hopping frequency $ \hat{f}_{hm}(={\hat{k}_{hm}B}/{K}) $. Comparing $ \{\hat{f}_{hm}~\forall m\} $ with the FHCS constellations, the FHCS sub-symbol is demodulated. 

\textit{b) PSK sub-symbol:} Using the identified $ k_{hm} $ and the estimations of channel parameters, we can estimate $ {\varpi}_{hm} $ based on (\ref{eq: DFT of com signal m-th frequency}), as given by $ \hat{\varpi}_{hm} = \angle\left( Y_{hm}^{(d^*)}\hat{\tilde{\beta}}^{\dagger}e^{\mj\frac{2\pi m \hat{u}}{M}}e^{-\mj \measuredangle\omega_{{\eta}} \hat{k}_{hm}  } \right) $, where $ ()^{\dagger} $ takes conjugate.
Comparing $ \hat{\varpi}_{hm} $ with the PSK constellation, the PSK sub-symbol is demodulated.

	\section{Extensions of Proposed Methods}\label{sec: extensions to NLOS and multi-antenna scenarios}
	In this section, we first extend the proposed methods to multi-path scenarios. We then discuss the potential extensions to multi-antenna receivers and the scenarios in the presence of interference. 	
	The two estimators, CAE and CRE, can be extended to flat Rician fading channels, where the radar-transmitted signal arrives at the communication receiver through several non-LoS (NLoS) paths in addition to the LoS. By considering a quasi-static flat-fading channel, the delay spread can be confined within a radar snapshot \cite{DFRC_ChnnlEst_FanLiu_howMuchInfo}.
	Thus, the waveform orthogonality defined in (\ref{eq: fphm ne fphm'}) is preserved, and the method developed in Section \ref{subsec: novel waveform and HF estimation} can still be used to estimate the hopping frequencies at the communication receiver, leading to the same result in (\ref{eq: k_ph^m}). 
	As in the LoS case, we use the frequency-domain signals, i.e., $ Y_m(l_{M-1-m}^*) $ in (\ref{eq: DFT of com signal m-th frequency}), to estimate communication channel. 
		
	Let $ p \in[0,P-1] $ denote path index. The multi-path version of (\ref{eq: DFT of com signal m-th frequency}) can be given by
	\begin{align} \label{eq: DFT of com signal m-th frequency multipath}
		Y_m  = Y(l_{M-1-m}^*)   
		= {\rho_m}   \omega_{\eta}^{\hat{k}_{m}}+\Xi_m,~\rho_m=\sum_{p=0}^{P-1}\tilde{\beta}_p  e^{-\mj\frac{2\pi m u_p}{M}}
	\end{align} 
	where variables are defined in the same way as in (\ref{eq: DFT of com signal m-th frequency}) and $\Xi(l_{M-1-m}^*) $ therein is shortened into $ \Xi_m$. 
	Compared with (\ref{eq: DFT of com signal m-th frequency}), the linear phase relation of the coefficients of $ \omega_{\eta}^{\hat{k}_{m}} $ has been destroyed in (\ref{eq: DFT of com signal m-th frequency multipath}) by NLoS components. 
	To apply the estimators CAE and CRE developed in Section \ref{sec: tau estimaiton}, we propose to estimate and suppress $ \rho_m $ first.

	From (\ref{eq: DFT of com signal m-th frequency multipath}), we observe that \textit{if the hopping frequency attached to the $ m $-th antenna is zero, i.e., $ \hat{k}_m=0 $, then $ Y_m =\rho_m + \Xi_m $ becomes an estimate of $ \rho_m $}. 
	Thus, $ \rho_0 $ can be estimated as $ Y_0 $, since $ k_0=0 $ is ensured in the sub-optimal hop sequence designed in Section \ref{subsec: design sub-optimal HFs}. 
	To estimate $ \rho_m~(m>0) $, we resort to the hops originally assigned for data communication; see Fig. \ref{fig: signal structure}(b). 
	In specific, we require that, for antenna $ m $ at hop $ (m+2) $, the hopping frequency is zero and no information bit is embedded, i.e., 
	\begin{align}\label{eq: khm=0, Fhm=1}
	k_{hm} = 0,F_{hm}=1,h=m+2,m=0,1,\cdots,M-1,
	\end{align}  
	where $ F_{hm}=1 $ denotes no information modulation; see (\ref{eq: comm-received signal continuous}).

	Referring to Fig. \ref{fig: signal structure}(b), the sampled hop $ (m+2) $ spans across hops $ (m+2) $ and $ (m+3) $. To avoid inter-hop interference, we propose to take an $ L $-dimensional yet $ \frac{L}{2} $-point DFT on the sampled hop $ (m+2) $. This leads to 
	\begin{align}\label{eq: DFT of hop h for rhom estimation}
	&Y_{h}(l) = 	\sum_{i=0}^{L/2-1} y_{h}(i)e^{-\mj\frac{2\pi il}{L}} = \sum_{m=0}^{M-1}  \rho_m e^{\mj \pi(L-1)}
		\omega_{{\eta}}^{k_{hm}}  \times \nonumber\\
		&~~~~~~ \frac{\sin\frac{\pi}{2}\left( \frac{k_{hm}BT}{K}+l \right)}{\sin\frac{\pi}{L}\left( \frac{k_{hm}BT}{K}+l \right)}e^{-\mj\frac{\pi(L-2)\left( \frac{k_{hm}BT}{K}+l \right) }{2L}} + \Xi(l),
	\end{align}
	where $ \rho_m $ is given in (\ref{eq: DFT of com signal m-th frequency multipath}), $ u $ and $ \omega_{{\eta}} $ are defined in (\ref{eq: intermediate variables}), and $ \Xi(l) $ denotes the DFT of the AWGN $ \xi(i) $. 
	Based on (\ref{eq: khm=0, Fhm=1}) and (\ref{eq: DFT of hop h for rhom estimation}), we have 
	\begin{align}\label{eq: Y_{m+2}(0)}
		& Y_{m+2}(0) = \frac{L}{2}\rho_m e^{\mj \pi(L-1)} +  \sum_{\substack{m'=0\\m'=\ne m}}^{M-1}  \rho_{m'} e^{\mj \pi(L-1)}
		\omega_{{\eta}}^{k_{hm'}}\times \nonumber\\
		& ~~~~~~~~~~~~~\frac{\sin \frac{k_{hm'}BT\pi}{2K}}{\sin \frac{k_{hm'}BT\pi}{LK}}e^{-\mj\frac{\pi(L-2) \frac{k_{hm'}BT}{K} }{2L}} + \Xi(0).
	\end{align}

We see from (\ref{eq: Y_{m+2}(0)}) that taking $ \frac{k_{hm'}BT}{2K} $ as an integer makes the sine function in the numerator become zero, hence avoiding inter-antenna interference. This can be achieved by configuring the hopping frequencies as
 	\begin{align} \label{eq: khm' for rhom estimation}
 		k_{hm'}=\left\{  
 		\begin{array}{ll}
 			\forall k\in [1,K-1]& \text{if }{BT/K}\text{ is even} \\
 			\forall k \in \{2,4,6,\cdots\}& \text{otherwise}
 		\end{array}
 		\right..
 	\end{align} 
 	Based on the above analyses and derivations, 
 	estimating $ \omega_{{\eta}} $ in Rician channels is summarized in the following proposition.
	
	\begin{proposition}
		\textit{Provided that (\ref{eq: khm=0, Fhm=1}) and (\ref{eq: khm' for rhom estimation}) are satisfied, CAE given in (\ref{eq: eta estimate based on bar mathcal M}) and CRE given in (\ref{eq: eta estimate based on breve mathcal M}) are capable of estimating $ \omega_{{\eta}} $ in Rician channels, by replacing $ Y_m~\forall m $ in (\ref{eq: breve{Y}}) with $ Y_m/\hat{\rho}_{m} $, where 
			\begin{align}
				\hat{\rho}_{m}=\left\{  
					\begin{array}{ll}
						Y_0\text{ based on (\ref{eq: DFT of com signal m-th frequency multipath})}& \text{at } m=0\\
						\frac{2}{L} Y_{m+2}(0) e^{-\mj\pi(L-1)}\text{ based on (\ref{eq: Y_{m+2}(0)})} & \text{elsewhere}
					\end{array}
				\right..\nonumber
			\end{align}	
	}
	\end{proposition}

With $ \omega_{{\eta}} $ estimated, two options are available for the remaining processing: (I) We can proceed to estimate $ u_0 $ and $ \beta_0 $ using the method developed in Section \ref{subsec: estimation of u and beta}, and then perform data communication as illustrated in Section \ref{sec: decoding  using estimated channels}; (II) We can suppress timing offset using the method provided in Section \ref{sec: decoding  using estimated channels}, and then divide $ Y_m $ by $ \hat{\rho}_m~\forall m $ to remove the impact of other channel parameters. 
The benefit of the first option is that, with the path AoD estimated, beamforming can be performed to enhance LoS and suppress NLoS paths, provided that multiple antennas are equipped. In contrast, when a single antenna is available, the second option can be more efficient in NLoS scenarios.

	Note that the proposed channel estimation methods can be extended to multi-antenna communication receivers. The signals received by different antennas only have phase differences that are caused by propagation delays and are constant over time. To this end, the proposed methods can be directly performed based on the signal received at any antenna. Moreover, accumulations across antennas can be carried out to improve estimation performance. 
	For example, 
	the signals can be {coherently} accumulated over antennas before performing CAE or CRE; see (\ref{eq: barY}). 
	
	Also note that the proposed channel estimation methods can work in the presence of interference, provided that there are enough 
	clean sub-bands without being interfered. From (\ref{eq: eta estimate based on bar mathcal M}) and (\ref{eq: eta estimate based on breve mathcal M}), we notice that the proposed CAE and CRE can work on a subset of $ \bar{\mathcal{M}} $ and $ \breve{{{\mathcal{M}}}} $, respectively. 
	To this end, we can identify the clean sub-bands and use them to perform CAE or CRE, as will be validated in simulations.
	In the case that there are not sufficient clean sub-bands, array-based interference nullifying can be resorted to. 
	This is left for future work.

\section{Simulation Results}\label{sec: simulations}
In this section, simulations are provided to validate the high accuracy of the proposed channel estimation methods. 
Unless otherwise specified, the FH-MIMO radar is configured as: $ M=10 $, $ K=20 $, $ B=100 $ MHz, $ f_{\mathrm{L}}=8 $ GHz, $ f_{\mathrm{s}}=2B $, $ T=0.8~\mu $s and {$ d=\frac{\lambda}{2} $}. 
Both LoS and Rician channels are simulated, where $ \eta\sim\mathcal{U}_{[0.05\mu\mathrm{s},0.35\mu\mathrm{s}]} $, $ \phi_0=20^{\circ} $, $ \phi_p\sim\mathcal{U}_{[-90^{\circ},90^{\circ}]}~(\forall p>0) $, $ \beta_0 = e^{\mj\angle \beta_0} $ with $ \angle\beta_0\sim\mathcal{U}_{[0^{\circ},360^{\circ}]} $, $ \beta_p~(\forall p>0)\sim\mathcal{CN}(0,-5\mathrm{~dB}) $ (the Rician factor is $ 5 $ dB), and $ P=4 $ NLoS paths are added for Rician channels. By taking $ p=0 $, LoS channels are obtained.  
Here, $ \mathcal{U}_{[\cdot,\cdot]} $ stands for the uniform distribution in the subscript region. Based on the above parameters, the sub-optimal \HF~sequence can be calculated as in Section \ref{subsec: design sub-optimal HFs}, leading to 
\[\text{sub-optimal: }\mathbf{f}^*=f_{\mathrm{L}}+[0,1,3,4,6,7,9,10,17,19]^{\mathrm{T}}\times {B}/{K}.\]
In addition, exploiting (\ref{eq: barY}) and (\ref{eq: bar mathcal M}), we can identify the \HF~sequence satisfying $ \bar{M}=(M-2) $, i.e., 
$ \text{CAE: }\bar{\mathbf{f}} = f_{\mathrm{L}}+[0,1,3,4,6,7,9,10,12,13]^{\mathrm{T}}\times {B}/{K}. $
When applying $ \bar{\mathbf{f}} $, only CAE is applicable with its MSELB minimized. 
Similarly, substituting the above parameters into (\ref{eq: barY}) and (\ref{eq: breve mathcal M}), the \HF~sequence leading to the minimum $ \breve{\sigma}^2_{\eta} $ can be obtained, as give by 
$ \text{CRE: }\breve{\mathbf{f}}=f_{\mathrm{L}}+[0,1,2,3,4,5,6,7,17,19]^{\mathrm{T}}\times {B}/{K}. $
Note that $  \breve{\mathbf{f}} $ leads to $ \bar{\mathcal{M}}=\emptyset $, and hence only CRE is applicable.
The above three sequences of \HFs~are adopted and compared in the following simulations.

\begin{figure}[!t]
	\centerline{\includegraphics[width=80mm]{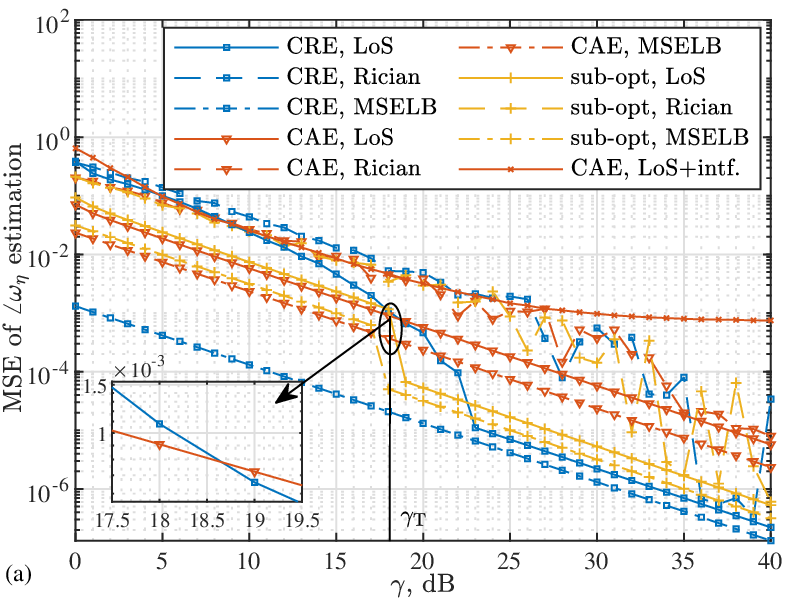}}
	\centerline{\includegraphics[width=80mm]{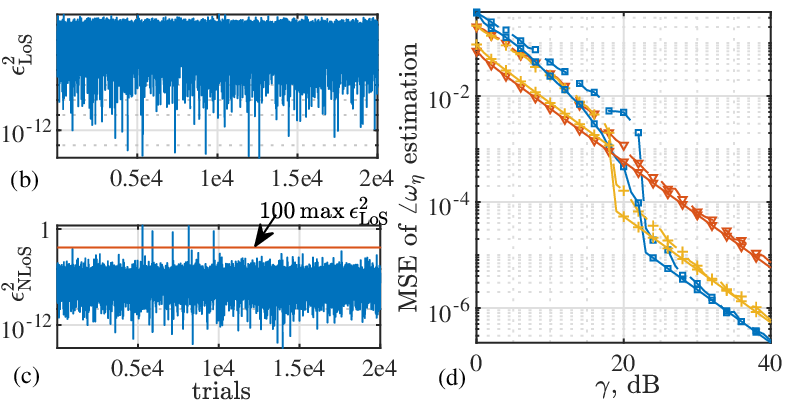}}
	\caption{{(a) MSE of $ \angle\omega_{\eta} $ estimation against $ \gamma(=|\beta|^2/\sigma_n^2) $, where ``intf.'' is short for interference, and $ \sigma_n^2 $ is the power of $ \xi(i) $; see (\ref{eq: communication signal p h digitized}) for $ \beta $ and $ \xi(i) $; (b) and (c) are obtained at $ \gamma=30 $ dB, where $ \epsilon_{\mathrm{LoS}}^2 $ and $ \epsilon_{\mathrm{NLoS}}^2 $ denote the squared estimation errors for LoS and NLoS scenarios, respectively; (d) MSE of $ \angle\omega_{\eta} $ estimation in NLoS scenarios with abnormal trials removed, where an abnormal trial has $ \epsilon_{\mathrm{NLoS}}^2>100\max\{\epsilon_{\mathrm{LoS}}^2\} $.}}
	\label{fig: mse eta estimation against snr}
\end{figure}

Fig. \ref{fig: mse eta estimation against snr}(a) plots the MSE of $ \angle\omega_{\eta} $ estimation against the received SNR $ \gamma $ at the communication receiver in LoS and Rician channels. 
We first analyze the estimation results for LoS channels. 
We see from the figure that CRE has a much better high-SNR performance than CAE and whereas CAE outperforms CRE in low SNR regions. 
This validates the analysis in Remark \ref{rmk: low snr comparison of estimates}. 
We also see that CRE and CAE are able to asymptotically approach their MSELBs
derived in Proposition \ref{pp: mselbs}. This validates the analysis in Appendix \ref{app: proof of MSELBs}.
By comparing CRE and CAE, the SNR threshold $ \gamma_{\mathrm{T}}=18 $ dB can be obtained from the zoomed-in turning point. As analyzed in Remark \ref{rmk: low snr comparison of estimates}, we perform CAE and CRE below and above $ \gamma_{\mathrm{T}} $, respectively, meanwhile exploiting the sub-optimal $ \mathbf{f}^* $. We see from Fig. \ref{fig: mse eta estimation against snr} that the sub-optimal $ \mathbf{f}^* $ provides the sub-optimal estimation accuracy in the whole SNR regions.
Nevertheless, we see that the sub-optimal $ \mathbf{f}^* $ improves the estimation accuracy obviously over CRE in low SNR regions $ (\gamma\le \gamma_{\mathrm{T}}) $, and substantially outperforms CAE in high SNR regions $ (\gamma> \gamma_{\mathrm{T}}) $. It is noteworthy that the improvement achieved by the sub-optimal $ \mathbf{f}^* $ across the whole SNR region is based on a single hop.

{For Rician channels, we see from Fig. \ref{fig: mse eta estimation against snr}(a) that the estimation performance improves with $ \gamma $ in overall. This validates the effective extension of the proposed methods to multi-path channels, as elaborated on in Section \ref{sec: extensions to NLOS and multi-antenna scenarios}. We also see oscillations in the MSE results, particularly in high SNR regions. The oscillations are caused by few trials whose estimation errors are abnormally larger than the overall MSE. (Note that the few abnormal estimations are caused by signal canceling among multiple paths in Rician channels.) This can be validated by jointly observing Figs. \ref{fig: mse eta estimation against snr}(b) and \ref{fig: mse eta estimation against snr}(c). By removing the abnormal trials, the estimation performance under Rician channels approach that under LoS channels, as demonstrated in Fig. \ref{fig: mse eta estimation against snr}(d). } 

{Fig. \ref{fig: mse eta estimation against snr}(a) also shows the robustness of the proposed CAE against interference, where we consider a $ -5 $ dB interference signal from another non-synchronized FH-MIMO radar. The interference radar is configured the same as the target radar except that its hopping frequencies are randomly taken (leaving the first three sub-bands used by CAE uninterfered). We see from Fig. \ref{fig: mse eta estimation against snr}(a) that CAE is robust to interference particularly in low SNR regions. This validates the conditional robustness of the proposed method to interference, as discussed in Section \ref{sec: extensions to NLOS and multi-antenna scenarios}. We also see that, unlike in interference-free scenarios, the MSE of CAE converges to about $ 9\times 10^{-4} $ as SNR increases. As expected, interference, rather than AWGN, dominates the estimation performance at high SNRs.  
}

\begin{figure}[!t]
	\centerline{\includegraphics[width=80mm]{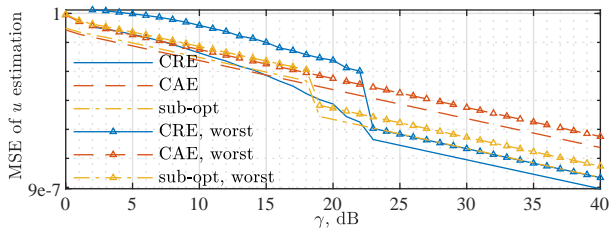}}
	\caption{MSE of $ u $ estimation against $ \gamma $, where the $ \angle\omega_{\eta} $ estimations from Fig. \ref{fig: mse eta estimation against snr} are used to calculate $ Z_m $ in (\ref{eq: tilde Z}) (as required for $ u $ estimation). The worst case adds the mean and standard deviation of squared estimation errors.}
	\label{fig: mse u estimation against snr}
\end{figure}

Fig. \ref{fig: mse u estimation against snr} observes the $ u $ estimation accuracy against $ \gamma $, where the $ \angle\omega_{\eta} $ estimations obtained from Fig. \ref{fig: mse eta estimation against snr} are used for calculating $ {Z}_m $ in (\ref{eq: tilde Z}). We see that the $ u $ estimation accuracy is closely dependent on the $ \angle\omega_{\eta} $ estimations. 
{In particular, we see a fast decay from $ 22 $ dB to $ 23 $ dB, when the $ \angle\omega_{{\eta}} $ estimates obtained by CRE are used for suppressing the timing offset. This is because the performance of CRE improves by an order of magnitude over the same SNR region; see Fig. \ref{fig: mse eta estimation against snr}(a).} We also see that the sub-optimal $ \mathbf{f}^* $ enables the MSE of $ u $ estimation to outperform those achieved by CRE and CAE in low and high SNR regions, respectively. 
On one hand, this validates the superiority of the proposed sub-optimal $ \mathbf{f}^* $ over $ \bar{\mathbf{f}} $ and $ \breve{\mathbf{f}} $, as consistent with Fig. \ref{fig: mse eta estimation against snr};
 and on the other hand, this demonstrates the robustness of the proposed $ u $ estimation to the estimation error of CAE and CRE, even in low SNR regions.

\begin{figure}[!t]
	\centerline{\includegraphics[width=80mm]{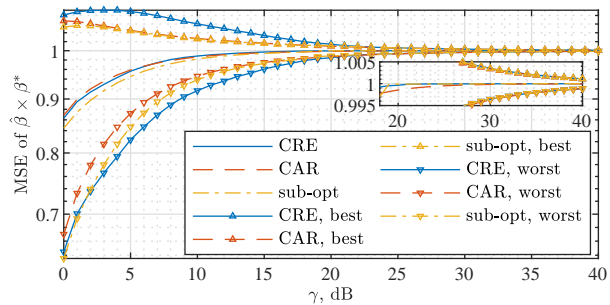}}
	\caption{MSE of $ \hat{\tilde{\beta}}\times \tilde{\beta}^* $ against $ \gamma $, where the $ u $ estimations in Fig. \ref{fig: mse u estimation against snr} are used to estimate $ \hat{\tilde{\beta}} $ as given in (\ref{eq: hat beta}) and $ \tilde{\beta}^* $ is the conjugate of $ \tilde{\beta} $. The best case subtracts the standard deviation of squared estimation errors from the MSE.}
	\label{fig: mse beta estimation against snr}
\end{figure}

Fig. \ref{fig: mse beta estimation against snr} plots the MSE of $ \hat{\tilde{\beta}}\times \tilde{\beta}^* $ against $ \gamma $, where the $ u $ estimations obtained in Fig. \ref{fig: mse u estimation against snr} are used in (\ref{eq: hat beta}). We see that, owing to the high accuracy of $ \angle\omega_{\eta} $ and $ u $, the estimate
$ \hat{\tilde{\beta}} $ is very close to the true value. 
From the zoomed-in sub-figure, we see that at $ \gamma=20 $ dB the MSE of $ \hat{\tilde{\beta}}\times \tilde{\beta}^* $ approaches to $ 1 $ with an error of less than $ 0.001 $. 
We also see that, due to the different performance index from that used in Figs. \ref{fig: mse eta estimation against snr} and \ref{fig: mse u estimation against snr}, the MSEs achieved by the three estimators are not as differentiable as they are in the previous two figures. Nevertheless, from the deviations of their squared errors, we see that CAE and the sub-optimal $ \mathbf{f}^* $ can produce better $ \tilde{\beta} $ estimations compared with CRE (particularly in low SNR regions).

\begin{figure}[!t]
	\centerline{\includegraphics[width=80mm]{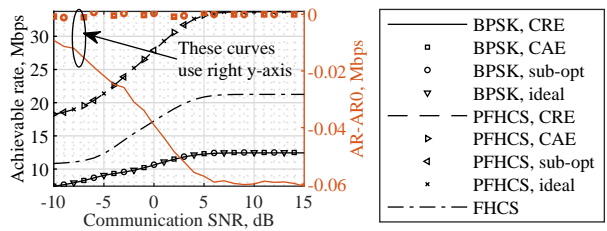}}
	\caption{Achievable date rate versus communication SNR, where AR0 is obtained based on the ideal channels.}
	\label{fig: rate versus snr}
\end{figure}

We proceed to demonstrate the efficacy of applying the channel estimations for data communications in FH-MIMO DFRC. Two existing constellations BPSK \cite{FH_MIMO_Radar2019_RadarConf} and FHCS \cite{DFRC_FHcodeSel2018}, and the new constellation PFHCS are evaluated based on the ideal and estimated channels. 
For fair comparison with BPSK \cite{FH_MIMO_Radar2019_RadarConf} and FHCS \cite{DFRC_FHcodeSel2018}, channel coding is not considered here.
Fig. \ref{fig: rate versus snr} illustrates the achievable data rate against the communication SNR. The estimations obtained at $ \gamma=15 $ dB from Figs. \ref{fig: mse eta estimation against snr}, \ref{fig: mse u estimation against snr} and \ref{fig: mse beta estimation against snr} are used to perform data communications. We see that the new constellation PFHCS improves the data rate substantially over BPSK and FHCS. In particular, the converging data rate of PFHCS is $ 170\%(=\frac{33.75-12.5}{12.5}) $ and $ 58.82\%(=\frac{33.75-21.25}{21.25}) $ higher than that of BPSK and FHCS, respectively. We also see that the estimated parameters enable the data rate to tightly approach that corresponds to the ideally known parameters. 
From the right $ y $-axis, we see that the sub-optimal estimator and CAE produce much smaller achievable rate difference across the whole communication SNR region, compared with CRE. This is consistent with the channel estimation accuracy shown in Figs. \ref{fig: mse eta estimation against snr}, \ref{fig: mse u estimation against snr} and \ref{fig: mse beta estimation against snr}.

\begin{figure}[!t]
	\centerline{\includegraphics[width=80mm]{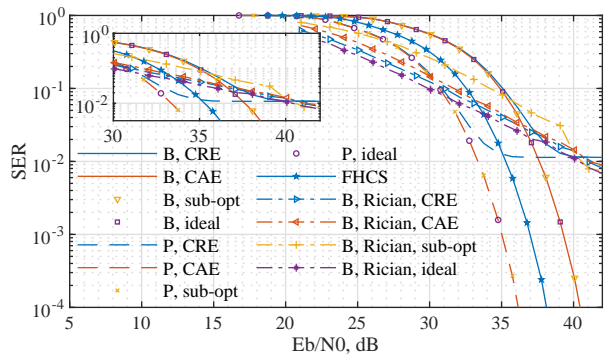}}
	\caption{SER versus Eb/N0, where Eb/N0 is the energy per bit to noise power density ratio and is calculated as $ L\gamma_{\mathrm{com}}BT/E $, where $ \gamma_{\mathrm{com}} $ denotes communication SNR and $ E $ is the number of bits conveyed per radar hop. In the legend, B and P are short for BPSK and PFHCS, respectively.}
	\label{fig: ser against snr}
\end{figure}

Fig. \ref{fig: ser against snr} compares the SERs achieved based on the ideal timing offset and channel parameters and the estimated ones, where the estimations obtained at $ \gamma=15 $ dB from Figs. \ref{fig: mse eta estimation against snr}, \ref{fig: mse u estimation against snr} and \ref{fig: mse beta estimation against snr} are used.
We see that due to the larger number of symbol bits of the new constellation PFHCS, its SER versus Eb/N0 is improved substantially, compared with that achieved by BPSK and FHCS. 
We also see that
the channel estimation error incurred by CRE makes the SERs of BPSK and PFHCS converge to $ 10^{-2} $, whereas the smaller estimation error of the sub-optimal estimator and CAE produces the continuously decreasing SERs against Eb/N0.
{Interestingly, we see from Fig. \ref{fig: ser against snr} that using channel estimations obtained under Rician channels can achieve better SER performance, compared with using the estimations under LoS channels. The main reason is that multiple paths can constructively enhance LoS signals and improve the detecting probability of identifying hopping frequencies; c.f., the destructive signal canceling leading to the abnormal $ \omega_{{\eta}} $ estimations in Fig. \ref{fig: mse eta estimation against snr}(c).}

\begin{figure}[!t]
	\centerline{\includegraphics[width=80mm]{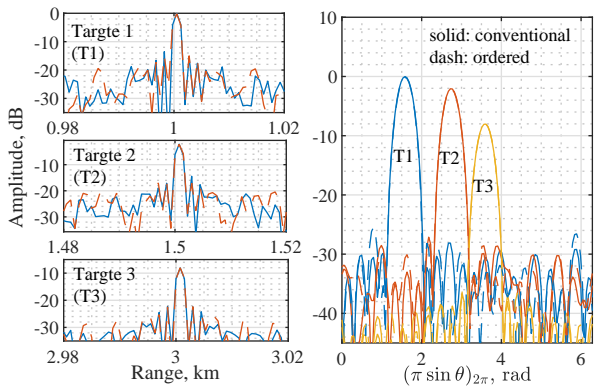}}
	\caption{Range and angle cuts when detecting radar targets. The range cuts are matched filter results with the propagation phases best compensated; similarly, the angle cuts are the DTFT over the extended receiver array at the radar range; see Steps 2\&3, Appendix \ref{app: radar signal processing}
		Here, $ (\cdot)_{2\pi} $ denotes modulo-$ 2\pi $.}
	\label{fig: radar range angle profile}
\end{figure}

\begin{figure}[!t]
	\centerline{\includegraphics[width=80mm]{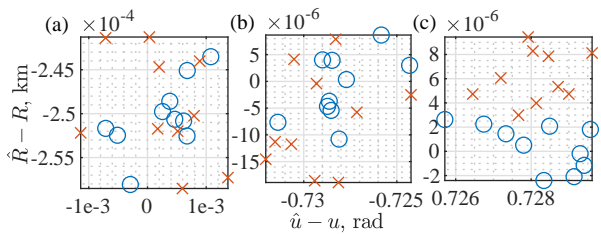}}
	\caption{Target detection results from $ 10 $ independent runs, where circle and cross markers denote results obtained using the (conventional) non-ordered and (proposed) ordered hopping frequencies, respectively. 
	}
	\label{fig: radar cfar}
\end{figure}

{Last but not least, we illustrate the impact of 
the proposed re-ordering of hopping frequencies per hop, referred to as waveform ordering below for brevity, on radar detection. 
Three targets are simulated. Their ranges are $ 1 $ km, $ 1.5 $ km and $ 3 $ km, and their directions are $ 30^{\circ} $, $ 60^{\circ} $ and $ -60^{\circ} $. Since Doppler estimation is not affected by waveform ordering \cite{Amb_FH_MIMO2008TSP}, we take a zero Doppler. The signal processing steps provided in Appendix \ref{app: radar signal processing} are used for target detection. Note that, for communication purpose, the first two hops per pulse are moderately adjusted to use the same hopping frequency sequence, i.e., the sub-optimal $ \mathbf{f}^* $; see the signal frame structure given in Fig. \ref{fig: signal structure}(b). The proposed waveform ordering is performed each hop. Moreover, BPSK phases are embedded for all antennas in the $ (H-2) $ hops assigned for data communications. The information bits are randomly generated per run. 
}

{
Fig. \ref{fig: radar range angle profile} plots the range cuts $ |X_p(\tau,u_r)|~(r=1,2,3) $ and the angle cuts $ |X_p(\tau_r,u)|~(r=1,2,3) $ of $ |X_p(\tau,u)| $ which is obtained in Step 3, Appendix \ref{app: radar signal processing}.
We see that the proposed waveform ordering does not affect the mainlobes of range and angle cuts and only incurs negligible changes to sidelobes. This is consistent with Proposition \ref{pp: range ambiguity}. The range and angle detecting results are further provided in Fig. \ref{fig: radar cfar}. 
We see again that the proposed waveform ordering produces similar detecting results to the conventional radar waveform. Confirmed by Figs. \ref{fig: radar range angle profile} and \ref{fig: radar cfar} and the previous results, 
we can conclude that our proposed scheme provides a promising solution to practical applications of FH-MIMO DFRC. 
}

\section{Conclusion and Future Work}\label{sec: conclusion}
This paper develops accurate methods to estimate timing offset and channel for FH-MIMO DFRC. 
This is achieved by a novel FH-MIMO radar waveform which enables a communication receiver to estimate, rather than acquiring from radar, the hopping frequency associated with each radar transmitter antenna.
This is also accomplished by two new estimators for timing offset suited for different \HF~sequences and an accurate channel estimation method. 
This is further fulfilled by an effective extension of the proposed methods to multi-path scenarios.
Simulation results validate the high accuracy of the proposed estimation methods and the high communication performance attained using estimated channels.
{As a future work, interference suppression based on antenna arrays will be studied for FH-MIMO DFRC.}

\section{Appendix}

\subsection{FH-MIMO Radar Signal Processing}\label{app: radar signal processing}

Based on (\ref{eq: transmitted signal p h m}), the $ m $-th radar antenna transmits $ s_m(t) = \sum_{h=0}^{H-1}s_{hm}(t)g_{{T}}(t-h{T}) $,
where $ g_{{T}}(t) $ is the step function taking unit one for $ 0\le t\le {T} $ and $ 0 $ elsewhere.
	Assume that there are $ N_{\mathrm{RT}} $ radar targets. In a co-located MIMO radar, the transmitter and receiver arrays see the $ r $-th target from the same direction, denoted by $ \theta_r $. Let $ \mathbf{a}(\theta_r) $ and $ \mathbf{b}(\theta_r) $ denote the array steering vectors of the radar transmitter and receiver arrays, respectively. 
	Assuming uniform linear array, the antenna spacing is $ \frac{\lambda}{2} $ and $ \frac{M\lambda}{2} $ for transmitter and receiver, respectively, where $ \lambda $ is radar wavelength. The $ m $-th element of $ \mathbf{a}(\theta_r) $ is $ e^{-\mj m\pi\sin\theta_r} $, 
	and the $ n $-th element of $ \mathbf{b}(\theta_r) $ is $ e^{-\mj nM\pi\sin\theta_r} $. 
	Given the round-trip propagation delay of the $ r $-th radar target, $ \tau_r $, and the target reflection coefficient, $ \alpha_r $, the signals received by antenna $ n $ 
	at PRI $ q $ is 
	\begin{align} %
	{x}_{qn}(t)=\sum_{r=0}^{N_{\mathrm{RT}}-1}
	\sum_{m=0}^{M-1}\alpha_r s_m(t-\tau_r)e^{\mj 2\pi f_{\mathrm{D}r}qT_{\mathrm{p}}}e^{-\mj (m+nM) u_r}, \nonumber   %
	\end{align} 
	where $ u_r=\pi\sin\theta_r $, $ f_{\mathrm{D}r} $ is the Doppler frequency of target $ r $ and $ T_{\mathrm{p}} $ denotes the PRI. 
	Due to the short length of radar pulse, $ f_{\mathrm{D}r} $ is approximately constant within a pulse \cite{Amb_FH_MIMO2008TSP}. 
	For notational simplicity, the noise term is dropped here.  
	Following \cite{Amb_FH_MIMO2008TSP,book_JianLi2008mimo}, the receiving processing of an FH-MIMO radar can be carried out as follows. 
	
	\textit{Step 1}, use matched filters to process the received signals, where the filter coefficients are the conjugate of transmitted signals. Employing the $ m' $-th $ (m'\in [0,M-1]) $ transmitted signal $ s_{m'}(t) $ as the matched filter coefficient, filtering $ {x}_{qn}(t) $ yields $ \tilde{x}_{qnm'}(\tau) = \sum_{r=0}^{N_{\mathrm{RT}}-1}
	\sum_{m=0}^{M-1}\alpha_r g_{mm'}(\tau-\tau_r)e^{\mj 2\pi f_{\mathrm{D}r}q T_{\mathrm{p}}}e^{-\mj (m+nM) u_r} $,
	where $ g_{mm'}(\tau)=\int_{\infty}^{\infty} s_m(t)s_{m'}^*(t-\tau) \mathrm{d}\tau $ \cite{Amb_FH_MIMO2008TSP}.

	\textit{Step 2}, take the discrete time Fourier transform (DTFT) of $ \tilde{x}_{q\tilde{n}}(\tau)  $ in terms of $ \tilde{n}=(m'+nM)=0,1,\cdots,NM-1 $, leading to $ X_q(\tau,u) = \sum_{\tilde{n}=0}^{NM-1} \tilde{x}_{q\tilde{n}}(\tau) e^{-\mj \tilde{n} u} $. 
	
	\textit{Step 3}, perform the constant false alarm rate (CFAR) detection on $ |X_q(\tau,u)| $ to detect targets and estimate their ranges and directions. 
	Refer to \cite[Ch. 9, 10]{book_JianLi2008mimo} for details on CFAR.

	\textit{Step 4}, perform moving target detection \cite[Ch. 9]{book_JianLi2008mimo} to estimate Doppler frequencies.

\subsection{Proof of Proposition \ref{pp: range ambiguity}}\label{app: proof on invariable range ambiguity function}
The proof can be established from analyzing the range ambiguity function of the FH-MIMO radar. 
Based on \cite[Eq. (27)]{Amb_FH_MIMO2008TSP}, the range ambiguity function of the radar, denoted by $ R(\tau) $, can be expressed as
\begin{align}\label{eq: chi(tau)}
R(\tau) = \left| \sum_{m=0}^{M-1}\sum_{m'=0}^{M-1}\sum_{h,h'=0}^{H-1} \underbrace{\chi(\tilde{\tau},{\nu})e^{\mj2\pi {\nu}{hT}}}_{\mathcal{B}} \underbrace{e^{\mj2\pi f_{h'm'}\tau}}_{\mathcal{D}} \right|,
\end{align}
where $ \tilde{\tau} = \tau-{T(h'-h)} $, $ {\nu}=f_{hm} - f_{h'm'} $ and $ \chi(x,y) $ is the ambiguity function of a standard rectangular pulse with $ x $ and $ y $ spanning range and Doppler domains, respectively. 
According to \cite[Eq. (26)]{Amb_FH_MIMO2008TSP}, we have $ \chi(x,y) = \Big(T-|x|\Big)\mathcal{S}\Big(y\big(T-|x|\big)\Big)e^{\mj\pi y(x+T)},~\mathrm{if}~|x|<T $; and otherwise $ \chi(x,y)=0 $, where $ \mathcal{S}(\alpha)=\frac{\sin(\pi\alpha)}{\pi\alpha} $ is the sinc function. 

As \HFs~are independently selected across hops, any change of $ f_{hm} $ at hop $ h $ has no impact on $ f_{h'm'} $ at hop $ h' $, and vice versa. Therefore, we can claim that the set of combinations of $ \big({\nu},f_{h'm'}\big) $ remain the same given any ordering of \HFs~at hops $ h $ and $ h' $. The underlying principle is that the overall combinations of $ \big({\nu},f_{h'm'}\big) $ are independent of element orderings \cite{book_combinatorics2010}.  

Moreover, we see from (\ref{eq: chi(tau)}) that the combinations of $ (\mathcal{B},\mathcal{D}) $ are uniquely determined by the combinations of $ \big({\nu},f_{h'm'}\big) $, since the other two parameters, $ \tilde{\tau} $ and $ \tau $, are independent of $ m$ or $m' $. 
Given the independence of the set, $ \{\big({\nu},f_{h'm'}\big)~\forall h,h',m,m' \} $, on the ordering of \HFs, we conclude that the range ambiguity function, $ R(\tau) $, is unaffected by the reordering introduced in (\ref{eq: new waveform}).

\subsection{Proof of Proposition \ref{pp: mselbs}}\label{app: proof of MSELBs}
The proof is established by first proving that the estimators proposed in (\ref{eq: eta estimate based on bar mathcal M}) and (\ref{eq: eta estimate based on breve mathcal M}) are MLEs, and then evaluating the estimation SNRs for the two estimators to derive their MSELBs.  
Let $ \Xi_m $ denote the noise term $ \Xi(l_{M-1-m}^*) $ in (\ref{eq: DFT of com signal m-th frequency}).
Here, $ {\Xi}_m $, as the DFT of the AWGN $ \xi(i) $, is still an AWGN; refer to (\ref{eq: DFT of digitized communication signal}). (The underlying principle is that linear calculations involved in DFT do not change the statistic distribution of AWGNs \cite{book_oppenheim1999discrete}.)
Substituting (\ref{eq: DFT of com signal m-th frequency}) into (\ref{eq: breve{Y}}), the noise term added to $ \breve{Y}_m $ can be given by 
\begin{align}\label{eq: breve Xi}
& \breve{\Xi}_m = \frac{{Y}_{m}+{\Xi}_m}{{Y}_{m+1} + {\Xi}_{m+1}} - \breve{Y}_m= \frac{\breve{Y}_m + \frac{{\Xi}_m}{{Y}_{m+1}}  }{1+\frac{{\Xi}_{m+1}}{{Y}_{m+1}}}  - \breve{Y}_m \nonumber\\
&\approx  \breve{Y}_m \left( 1-\frac{{\Xi}_{m+1}}{{Y}_{m+1}} \right) +  \frac{{\Xi}_m}{{Y}_{m+1}}\left( 1-\frac{{\Xi}_{m+1}}{{Y}_{m+1}} \right) - \breve{Y}_m,\nonumber\\
& = -\frac{\breve{Y}_m {\Xi}_{m+1}}{{Y}_{m+1}} + \frac{{\Xi}_m}{{Y}_{m+1}} - \frac{{\Xi}_m}{{Y}_{m+1}}\frac{{\Xi}_{m+1}}{{Y}_{m+1}}
\end{align}
where the approximation is based on the Taylor series of $ \frac{1}{1+x}=1-x~\forall x\ll 1 $ \cite{FreqEst2019TCOM_Serbes}.

Similarly, by substituting (\ref{eq: breve{Y}}) and (\ref{eq: breve Xi}) into (\ref{eq: barY}), the noise term added to $ \bar{Y}_m $ becomes
\begin{align}\label{eq: bar Xi}
&\bar{\Xi}_m = \frac{\breve{Y}_{m} + \breve{\Xi}_m}{\breve{Y}_{m+1} + \breve{\Xi}_{m+1}} -\bar{Y}_m = 
\frac{\bar{Y}_m + \frac{\breve{\Xi}_m}{\breve{Y}_{m+1}}  }{1+\frac{\breve{\Xi}_{m+1}}{\breve{Y}_{m+1}}}  - \bar{Y}_m   \nonumber\\
&~~~~\approx \frac{\bar{Y}_m {\Xi}_{m+2}}{{Y}_{m+2}}
-\frac{\bar{Y}_m {\Xi}_{m+1}}{\breve{Y}_{m+1}{Y}_{m+2}}
- \frac{\breve{Y}_m {\Xi}_{m+1}}{\breve{Y}_{m+1}{Y}_{m+1}}  +  \frac{{\Xi}_{m}}{\breve{Y}_{m+1}{Y}_{m+1}}    \nonumber\\
& 
~~~~
+  \frac{\bar{Y}_m {\Xi}_{m+1}{\Xi}_{m+2}}{\breve{Y}_{m+1}{Y}_{m+2}^2}   - \frac{ {\Xi}_{m+1}{\Xi}_{m}}{\breve{Y}_{m+1}{Y}_{m+1}^2} - \frac{\breve{\Xi}_m}{\breve{Y}_{m+1}}\frac{\breve{\Xi}_{m+1}}{\breve{Y}_{m+1}}
\end{align}
where we have used the same mathematical manipulations as those applied in (\ref{eq: breve Xi}). From the most RHS of (\ref{eq: bar Xi}), we see that only the first four terms dominate the statistical distribution of $ \bar{\Xi}_m $, since the other terms have the products of at least two independent AWGNs. Thus, we obtain that the additive noise to $ \bar{Y}_m $ approaches to an AWGN in high SNR regions.

In the background of $ \bar{\Xi}_{\tilde{m}}~(\forall \tilde{m}\in \mathcal{M}\subseteq \{\bar{Y}_m \forall m\}) $, $ \measuredangle \bar{Y}_{\tilde{m}} $ is an MLE of the phase of $ \bar{Y}_{\tilde{m}} $, since the angle estimate of a complex number corrupted by AWGN is an unbiased MLE \cite{book_JianLi2008mimo}. 
Accordingly, $ {\measuredangle\omega}_{\eta}(d_{\tilde{m}}) $ obtained in (\ref{eq: hat eta m}) is an MLE due to the linear relation between $ {\measuredangle\omega}_{\eta}(d_{\tilde{m}}) $ and $ \measuredangle \bar{Y}_{\tilde{m}} $. 
By comparing (\ref{eq: hat eta m}) and (\ref{eq: eta estimate based on bar mathcal M}), we also see that $ {\measuredangle}\bar{\omega}_{\eta} $ is an MLE, since the summation does not change the AWGN distribution. Moreover, by substituting (\ref{eq: hat eta m}) into (\ref{eq: eta estimate based on breve mathcal M}), we further see that $ {\measuredangle}\breve{\omega}_{\eta} $ is an MLE, since $ {\measuredangle}\breve{\omega}_{\eta} $ is a constant-scaled sum of the MLEs $ {\measuredangle\omega}_{\eta}(d_{\breve{m}}) $.
Being MLEs, the high-SNR MSELB of $ \measuredangle\bar{Y}_{\tilde{m}} $, $ {\measuredangle}\bar{\omega}_{\eta} $ and $ {\measuredangle}\breve{\omega}_{\eta} $ can be approximated as the reciprocal of the doubled estimation SNRs \cite{CRLB_MLEphaseEstimator1997TCOM}. The estimation SNRs of $ {\measuredangle}\bar{\omega}_{\eta} $ and $ {\measuredangle}\breve{\omega}_{\eta} $ are calculated below.

We start by deriving the estimation SNR of $ \measuredangle\bar{Y}_{\tilde{m}} $, denoted by $ {\Upsilon} $, since $ {\Upsilon} $ is the basis of the estimation SNRs of $ {\measuredangle}\bar{\omega}_{\eta} $ and $ {\measuredangle}\breve{\omega}_{\eta} $; see (\ref{eq: eta estimate based on bar mathcal M}) and (\ref{eq: eta estimate based on breve mathcal M}). 
Seen from (\ref{eq: barY}), the signal power of $ \bar{Y}_{\tilde{m}} $ is unit one; hence we have $ {\Upsilon}=\frac{1}{\bar{\sigma}_n^2} $, where $ \bar{\sigma}_n^2 $ is the noise variance of $ \bar{\Xi}_m $. 
Based on (\ref{eq: breve{Y}}) and (\ref{eq: barY}), we notice that the second and third terms in (\ref{eq: bar Xi}) are identical. Therefore, we have $ \bar{\sigma}_n^2=\frac{6\tilde{\sigma}_n^2}{|\tilde{\beta}|^2} $, where $ \tilde{\sigma}_n^2 $ denotes the noise variance of $ {\Xi}_m~\forall m $ and $ |\tilde{\beta}|^2 $ is the power of the signal component in (\ref{eq: DFT of com signal m-th frequency}). Based on (\ref{eq: intermediate variables}), we have $ |\tilde{\beta}|^2=L^2|\beta|^2 $.
Based on (\ref{eq: DFT of digitized communication signal}) and (\ref{eq: DFT of com signal m-th frequency}), we obtain $ \tilde{\sigma}_n^2 = L{\sigma_n^2} $, where $ L $ is the number of DFT points, and $ \sigma_n^2 $ denotes the noise variance of the time-domain AWGN $ \xi(i) $; see (\ref{eq: communication signal p h digitized}). 
The above calculations lead to $ {\Upsilon}=\frac{L|\beta|^2}{6{\sigma}_n^2} $. 

As the summation in (\ref{eq: eta estimate based on bar mathcal M}) is a coherent accumulation, we have $ \bar{\Upsilon}=\bar{M}{\Upsilon}=\frac{\bar{M}L|\beta|^2}{6{\sigma}_n^2} $ with $ \bar{\Upsilon} $ denoting the estimation SNR for $ {\measuredangle}\bar{\omega}_{\eta} $. By calculating $ \frac{1}{2\bar{\Upsilon}} $, the MSELB of $ {\measuredangle}\bar{\omega}_{\eta} $ is achieved in (\ref{eq: mselb of bar mathcal M}). 	
Based on (\ref{eq: eta estimate based on breve mathcal M}), the variance of $ {\measuredangle}\breve{\omega}_{\eta} $ can be approximated as $ \mathbb{E}\left\{ \left( \frac{1}{\breve{M}} \sum_{\breve{m}\in\breve{\mathcal{M}}}{\measuredangle}\omega_{\eta}(d_{\breve{m}}^*)-\angle\omega_{\eta} \right)^2\right\} $. Here, $ \angle\omega_{\eta} $ is used as the mean of $ {\measuredangle}\breve{\omega}_{\eta} $, since $ {\measuredangle}\omega_{\eta}(d_{\breve{m}}^*) $ is unbiased in the sense of $ \mathbb{E}\{{\measuredangle}\omega_{\eta}(d_{\breve{m}}^*)\}=\angle\omega_{\eta} $ \cite{CRLB_MLEphaseEstimator1997TCOM}. By suppressing the cross-terms, the above variance is lower bounded by $ \frac{1}{\breve{M}^2} \sum_{\breve{m}\in\breve{\mathcal{M}}}$ $\mathbb{E}\left\{ \left( {\measuredangle}\omega_{\eta}(d_{\breve{m}}^*)-\angle\omega_{\eta} \right)^2\right\}  =  \frac{1}{\breve{M}^2} \sum_{\breve{m}\in\breve{\mathcal{M}}}\sigma_{\breve{m}}^2, $
where $ \sigma_{\breve{m}}^2 $ denotes the MSELB of $ {\measuredangle}\omega_{\eta}(d_{\breve{m}}^*) $. From (\ref{eq: eta estimate based on breve mathcal M}), we obtain $ \sigma_{\breve{m}}^2= \frac{1}{2{\Upsilon}\kappa_{\breve{m}}^2}  $, which gives (\ref{eq: mselb of breve mathcal M}).

\bibliographystyle{IEEEtran}
\bibliography{IEEEabrv,../ref/bib_JCAS.bib}

\end{document}